\documentclass[english,aps,reprint,twocolumn,superscriptaddress,longbibliography,showkeys,showpacs]{revtex4-2}
\usepackage[english]{babel}

\usepackage{amssymb}
\usepackage{graphicx,xcolor}
\usepackage[unicode,bookmarksnumbered,breaklinks,colorlinks,linktocpage,
citecolor=blue,linkcolor=darkred,urlcolor=darkblue]{hyperref}
\usepackage{bm}
\usepackage{hyperref}
\usepackage[fleqn]{amsmath}

\definecolor{darkblue}{rgb}{0.0, 0.0, 0.55}
\definecolor{darkgreen}{rgb}{0.0, 0.2, 0.13}
\definecolor{darkred}{rgb}{0.55, 0.0, 0.0}

\begin{document}

\title{Frustrated kagome-lattice bilayer quantum Heisenberg antiferromagnet}

\author{Dmytro~Yaremchuk} 
\email{yaremchuk@icmp.lviv.ua}
\affiliation{Institute for Condensed Matter Physics,
    National Academy of Sciences of Ukraine,
    Svientsitskii Street 1, 79011 L'viv, Ukraine}
\affiliation{Institute of Applied Mathematics and Fundamental Sciences, 
	L'viv Polytechnic National University, 79013 L'viv, Ukraine}

\author{Taras~Hutak}
\email{t.hutak@icmp.lviv.ua}
\affiliation{Institute for Condensed Matter Physics,
	National Academy of Sciences of Ukraine,
	Svientsitskii Street 1, 79011 L'viv, Ukraine}

\author{Vasyl~Baliha}
\affiliation{Institute for Condensed Matter Physics,
	National Academy of Sciences of Ukraine,
	Svientsitskii Street 1, 79011 L'viv, Ukraine}
\affiliation{Department for General Physics,
	Ivan Franko National University of L’viv, 
	Kyrylo and Mefodiy Street 8, 79005 L’viv, Ukraine}

\author{Taras~Krokhmalskii}
\affiliation{Institute for Condensed Matter Physics,
	National Academy of Sciences of Ukraine,
	Svientsitskii Street 1, 79011 L'viv, Ukraine}

\author{Oleg~Derzhko}
\email{derzhko@icmp.lviv.ua}
\affiliation{Institute for Condensed Matter Physics,
	National Academy of Sciences of Ukraine,
	Svientsitskii Street 1, 79011 L'viv, Ukraine}
\affiliation{Fakult\"{a}t f\"{u}r Physik, 
	Universit\"{a}t Bielefeld, 
	Postfach 100131, 33501 Bielefeld, Germany}
\affiliation{Professor Ivan Vakarchuk Department for Theoretical Physics,
	Ivan Franko National University of L’viv, 
	Drahomanov Street 12, 79005 L’viv, Ukraine} 

\author{J\"{u}rgen~Schnack}
\affiliation{Fakult\"{a}t f\"{u}r Physik, 
	Universit\"{a}t Bielefeld, 
	Postfach 100131, 33501 Bielefeld, Germany}

\author{Johannes~Richter}
\affiliation{Institut f\"{u}r Physik, Otto-von-Guericke-Universit\"{a}t Magdeburg,
	P.O. Box 4120, 39016 Magdeburg, Germany}

\date{\today}

\begin{abstract}
We consider the $S=1/2$ antiferromagnetic Heisenberg model on a frustrated kagome-lattice bilayer with strong nearest-neighbor interlayer coupling and examine its low-temperature magnetothermodynamics using a mapping onto a rhombi gas on the kagome lattice. Besides, we use finite-size numerics to illustrate the validity of the classical lattice-gas description. 
Among our findings there are i) the absence of an order-disorder phase transition and ii) the sensitivity of the specific heat at low temperatures to the shape of the system just below the saturation magnetic field even in the thermodynamic limit.
\end{abstract}


 
\maketitle

\section{Frustrated bilayer Heisenberg antiferromagnets}
\label{s1}

Two spins one-half entangled in a singlet state are in the heart of quantum mechanics. In the theory of spin lattices, singlets emerging due to the antiferromagnetic Heisenberg interaction lead to various fascinating valence-bond solid or valence-bond liquid quantum states. The resonating valence-bond state proposed by P.~W.~Anderson as the ground state of spin liquids is a famous example of the latter kind of states \cite{Savary2017}. Valence-bond solid states, on the other hand, imply a short-range pairing and the formation of localized static bonds obeying certain long-range order. On some lattices the antiferromagnetic Heisenberg interaction promotes the emergence of valence-bond crystal states as, for instance, in the Majumdar-Ghosh model \cite{Majumdar1969a,Majumdar1969b}. Among such lattices one may also mention frustrated bilayer lattices which have been dealt with in a number of earlier publications, see Refs.~\cite{Lin2000,Richter2006,Chen2010,Derzhko2010,Albuquerque2011,Oitmaa2012,Murakami2013,Tanaka2014,Alet2016,Krokhmalskii2017,Richter2018,Stapmanns2018,Strecka2018,Kurita2019,Fan2024}.

\begin{figure}
\includegraphics[width=\columnwidth]{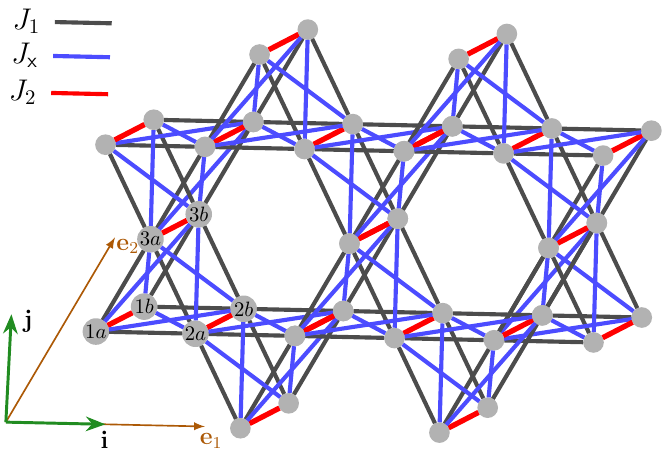}
\caption{Frustrated kagome-lattice bilayer. The lattice sites are denoted by ${\bf m}\alpha l$, where ${\bf m}=m_1{\bf e}_1+m_2{\bf e}_2$, $m_1,m_2$ are integers, ${\bf e}_1=2{\bf i}=(2,0)$, ${\bf e}_2={\bf i}+\sqrt{3}{\bf j}=(1,\sqrt{3})$ are the unit lattice vectors (here the triangle side has unit length), $\alpha=1, 2, 3$ denote the sites of the unit cell within a layer, and $l=a,b$ is the layer label.}
\label{f01}
\end{figure}

Frustrated bilayer systems consist of two identical layers, $a$ and $b$, with an intralayer interaction $J_1$, a nearest-neighbor interlayer interaction $J_2$, and a (frustrated) next-nearest-neighbor interlayer interaction $J_{\sf x}$, compare Fig.~\ref{f01} as an example.
If $J_1=J_{\sf x}=J$ one faces the fully frustrated case. 
For the fully frustrated case, a local singlet on a $J_2$ bond $(\vert\uparrow_a\downarrow_b\rangle - \vert\downarrow_b\uparrow_a\rangle)/\sqrt{2}$ in the environment of the remaining polarized spins is an eigenstate of the Hamiltonian.
Moreover, a product of such singlets surrounded by the polarized spins is also an eigenstate of the Hamiltonian.
Furthermore, if $J_2>J_{2c}(J)$ for some critical $J_{2c}(J)$, these states are the lowest-energy eigenstates and therefore may dominate the low-temperature magnetothermodynamics of the frustrated quantum spin systems in question. 

Since the localized eigenstates for the frustrated square-, honeycomb-, and triangular-lattice bilayers can be mapped onto spatial configurations of squares or hexagons on the square or hexagonal/triangular lattice, their contribution to thermodynamics is accounted for by means of classical statistical mechanics \cite{Derzhko2010,Krokhmalskii2017,Strecka2018}.
The most interesting prediction for these bilayers is an order-disorder phase transition related to singlet ordering:
Just below the saturation magnetic field $h_{\rm sat}$ the ground state is the two-fold (square and honeycomb bilayers) or three-fold (triangular bilayer) degenerate gapped localized-singlet crystal state, which corresponds to an ordered pattern of singlets respecting hard-core rules.
Such a (spontaneously chosen) ordered pattern persists up to some finite temperature $T_c(h)$ while further temperature increase drives the system into a disordered phase through a phase transition of Ising (square and honeycomb bilayers) or three-state Potts (triangular bilayer) universality class \cite{Derzhko2010,Krokhmalskii2017,Strecka2018}. 

Similarly, the localized eigenstates for the frustrated kagome-lattice bilayer can be mapped onto rhombi on the kagome lattice. However, since the kagome lattice can be covered by hard rhombi (nearest-neighbor exclusion) in a huge number of ways which grows exponentially with lattice size, one cannot expect an order-disorder phase transition pertaining to singlet ordering. This difference between the already studied frustrated bilayers and the not yet examined frustrated kagome-lattice bilayer inspired us to take a closer look at the latter in this publication.

In the present paper,
we consider the $S=1/2$ isotropic Heisenberg Hamiltonian  
\begin{eqnarray}
\label{01}
H=\sum_{\langle pq\rangle} J_{pq}{\mathbf s}_{p}\cdot{\bf s}_{q} -h\sum_{p} s_p^z,	
\end{eqnarray}
where the first (second) sum runs over the edges (vertices) of the frustrated kagome-lattice bilayer shown in Fig.~\ref{f01}, all exchange interactions are antiferromagnetic $J_{pq}>0$, and $h\ge 0$ is an external magnetic field. The total $S^z=\sum_ps^z_p$ commutes with $H$ and the subspaces with different good quantum numbers $S^z$ can be considered separately.  As said above, we distinguish the intralayer interactions $J_1$, the nearest-neighbor interlayer interactions $J_2$, and the next-nearest-neighbor interlayer interactions $J_{\sf x}$ (cf. black, red, and blue edges in Fig.~\ref{f01}). We set $(J_1+J_{\sf x})/2=J=1$ to fix the units.
The goal of the present study is to examine thoroughly the ground-state and finite-temperature properties of the frustrated kagome-lattice bilayer quantum Heisenberg antiferromagnet following the treatment of Refs.~\cite{Derzhko2010,Krokhmalskii2017,Strecka2018}.  

The remainder of this paper is organized as follows. 
We begin with numerics for finite-size systems, Section~\ref{s2}, focusing on the fully frustrated case $J_1=J_{\sf x}$.
Then we turn to analytical studies in Section~\ref{s3}.
We discuss one-magnon and many-magnon eigenstates as well as a mapping onto classical lattice models, and calculate the low-temperature magnetothermodynamics of the frustrated kagome bilayer.
In Section~\ref{s4}, we present an effective low-energy theory for a slightly violated fully frustrated condition $J_1\neq J_{\sf x}$.
Finally, we summarize our findings in Section~\ref{s5}.

\section{Numerical calculations for small finite-size lattices}
\label{s2}

In our numerical calculations,
we consider several finite-size frustrated kagome-lattice bilayers, in particular, with $N=36$ and $N=42$ (i.e., with the number of sites in one layer ${\cal N}=18$ and ${\cal N}=21$). Periodic boundary conditions are imposed.
Two shapes of the cluster with $N=36$ are determined by the edge vectors ${\bf a}_1=(6,0)=3{\bf e}_1$, ${\bf a}_2=(2,2\sqrt{3})=2{\bf e}_2$ ($N=36_a$) and ${\bf a}_1=(6,0)$, ${\bf a}_2=(0,2\sqrt{3})$ ($N=36_b$), see Appendix~\ref{aa}. The cluster with $N=42$ is determined by the edge vectors ${\bf a}_1=(7,-\sqrt{3})$, ${\bf a}_2=(0,2\sqrt{3})$.
For ideal flat-band geometry $J_1=J_{\sf x}=1$ we examine several $J_2=1,2,3,5$, see Secs.~\ref{s2}, \ref{s3}, and Appendix~\ref{aa}.
Small deviations from the flat band geometry are represented by the set $J_1=1.1$, $J_{\sf x}=0.9$, $J_2=5$, see Sec.~\ref{s4}.
We use full diagonalization for large $S^z$ or the finite-temperature Lanczos method \cite{Jaklic1994,Schnack2020} to calculate various ground-state characteristics (energy and degeneracy of low-lying levels, magnetization curve) and temperature dependencies of magnetization, susceptibility, entropy, specific heat.
Finite-size calculations play a role of ``experiments'', which are then explained by characterizing the low-energy eigenstates and mapping onto lattice gases.

\begin{figure}
\includegraphics[width=\columnwidth]{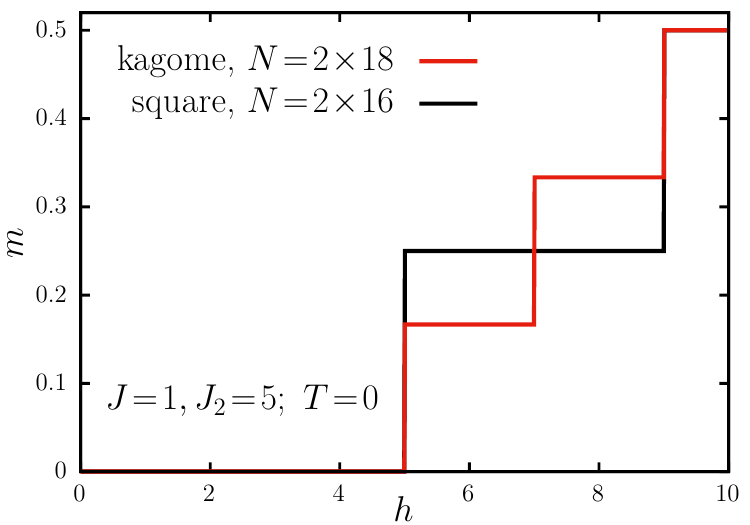}
\caption{Zero-temperature magnetization curve for the $S=1/2$ fully frustrated kagome-lattice bilayer with $J=1$, $J_2=5$, $N=36$ (red curve). Black curve corresponds to the square-lattice counterpart, $N=32$. There are no finite-size effects for this set of parameters.}
\label{f02}
\end{figure}

The typical ground-state magnetization curve for the large-$J_2$ regime shown in Fig.~\ref{f02} as red curve, has a 0-, a 1/3-, and a 2/3-magnetization  plateau for $0<h<h_1$, $h_1<h<h_2$, and $h_2<h<h_{\rm sat}$, respectively. Here $h_1=J_2$, $h_2=J_2+2J$, and $h_{\rm sat}=J_2+4J$. The plateau states are associated with the respective number of singlets in the unit cell. For example, the 2/3 plateau corresponds to one singlet and two polarized triplets, whereas the 1/3 plateau corresponds to two singlets and one polarized triplet within each unit cell. Interestingly, both these plateau states have a large degeneracy: For $S^z/(N/2)=2/3$, ${\cal W}_{\rm GS}=17,20,31$ for $N=36_a,36_b,42$, respectively.
The ground-state magnetization curve for the square-lattice bilayer, black curve in Fig.~\ref{f02}, exhibits only a 0 and a 1/2 plateau for $0<h<h_1$ and $h_1<h<h_{\rm sat}$. The former state corresponds to a crystal of singlets at all $J_2$ bonds. The latter state corresponds to a crystal state with checkerboard order of singlets which alternate with polarized triplets; such a state is only two-fold degenerate, see Refs.~\cite{Richter2006,Derzhko2010}.

More numerical results for the ground states and excitations are reported in Appendix~\ref{aa}. Most importantly, the data for low-energy eigenstates in the subspaces with $S^z=N/2-1,\ldots,N/3$ clearly illustrate the emergence of the large-$J_2$ regime: For $J_2>2J$, the low-energy eigenstates have easily understandable energy and degeneracy, see Sec.~\ref{s3} below. Moreover, the ground-state degeneracy ${\cal W}_{\rm GS}$ does not depend on $J_2$, and ${\cal W}_{\rm GS}$ for the 2/3 plateau state is rather large and grows with increasing $N$ (17, 20 or 31 for $N=36_a,36_b$ or $N=42$). For $J_2< 2J$ the ground state and the first excitated state are of completely different nature without regularly varying energies and large degeneracies.

Numerical results for finite temperatures are reported in Sec.~\ref{s3}. In particular, we present data for the temperature dependence of magnetization, susceptibility, entropy, and specific heat.
Moreover, we present some of such data at zero and finite temperatures for the case $J_1\neq J_{\sf x}$ in Sec.~\ref{s4}.

\section{Analytical and computer-aided calculations}
\label{s3}

\subsection{Flat bands and localized states}
\label{s3a}

\begin{figure}
\includegraphics[width=\columnwidth]{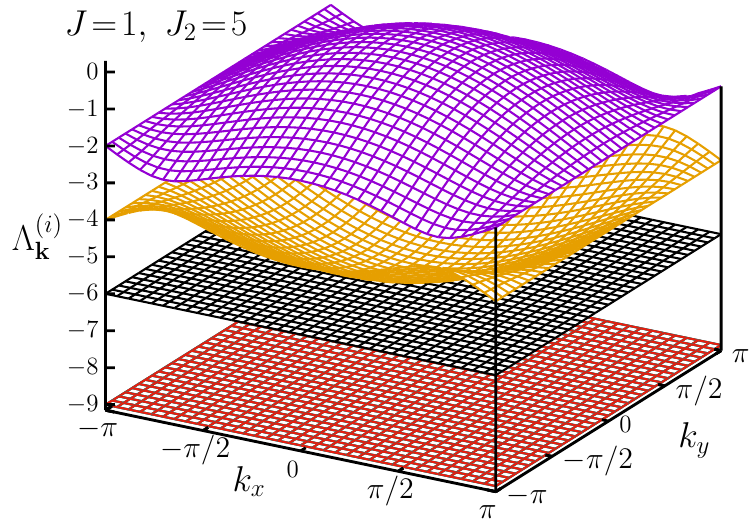}
\caption{One-magnon spectrum for $J=1$, $J_2=5$. The lowest-energy flat band is three-fold degenerate; setting $J_1=1.1\neq J_{\sf {x}}=0.9$ lifts the degeneracy and makes two of the three lowest-energy bands slightly dispersive, see Eq.~(\ref{05}).}
\label{f03}
\end{figure}

We begin this section with the one-magnon spectrum of the system at hand to show that this lattice supports completely dispersionless (flat) magnon bands.
To this end, we proceed as follows. 
First, set for a while $h=0$ in Eq.~(\ref{01}) and represent the Hamiltonian in the form: $H=\sum_{l=a,b}H_l+H_{ab}$, where $H_l$ is the kagome-layer Hamiltonian and $H_{ab}$ stands for the interlayer coupling. 
Second, for the one-magnon spectrum calculations, we have to replace ${\bf s}_{p}\cdot{\bf s}_{q}\to (s^+_{p}s^-_{q}+s^+_{q}s^-_{p})/2 -s^+_{p}s^-_{p}/2-s^+_{q}s^-_{q}/2+1/4$
(here the term $s^+_{p}s^-_{p}s^+_{q}s^-_{q}$ is omitted as irrelevant one).
Third, as usual, to convert from the ${\bf m}$-space to the ${\bf k}$-space, we introduce  
\begin{eqnarray}
\label{02}
s_{{\bf k}\alpha l}^{\pm}
=
\frac{1}{\sqrt{\cal{N}}}\sum_{\bf m}{\rm e}^{\mp{\rm i}{\bf k}\cdot{\bf m}} s_{{\bf m}\alpha l}^{\pm},
\nonumber\\
s_{{\bf m}\alpha l}^{\pm}
=
\frac{1}{\sqrt{\cal{N}}}\sum_{\bf k}{\rm e}^{\pm{\rm i}{\bf k}\cdot{\bf m}} s_{{\bf k}\alpha l}^{\pm},	
\end{eqnarray}
where ${\bf k}=(k_1,k_2)$ acquires ${\cal N}/3$ values: $k_{1,2}=2\pi z_{1,2}/{\cal L}_{1,2}$, $z_{1,2}$ obtains ${\cal L}_{1,2}$ integer values, ${\cal L}_1{\cal L}_2={\cal N}$, ${\cal N}=N/2$, and $k_1=k_x$, $k_2=(k_x-\sqrt{3}k_y)/2$.
Fourth, it is convenient to use matrix notations. As a result, the Hamiltonian (\ref{01}) in one-magnon space can be cast into
\begin{eqnarray}
\label{03}
H_{\rm 1m}
&=&
\left(J_1{+}J_{\sf x}{+}\frac{J_2}{4}\right){\cal N}
\nonumber\\
\!{+}\!&&\!\!\!\!\!\sum_{\bf k}
\left(
\begin{array}{cccccc}
\!s^{-}_{{\bf k}1a}\! & \!s^{-}_{{\bf k}2a}\! & \!s^{-}_{{\bf k}3a}\! & \!s^{-}_{{\bf k}1b}\! & \!s^{-}_{{\bf k}2b}\! & \!s^{-}_{{\bf k}3b}\!
\end{array}
\right)
\!{\bf H}\!
\left(\!
\begin{array}{c}
s^{+}_{{\bf k}1a} \\
s^{+}_{{\bf k}2a} \\
s^{+}_{{\bf k}3a} \\
s^{+}_{{\bf k}1b} \\
s^{+}_{{\bf k}2b} \\
s^{+}_{{\bf k}3b}
\end{array}
\!\right)\!,
\end{eqnarray}
where the first term is the ferromagnetic-state energy, 
and 
\begin{eqnarray}
\label{04}
{\bf H}
&=&
\left(
\begin{array}{cc}
{\bf A} & {\bf B} \\
{\bf B} & {\bf A}
\end{array}
\right)\ ,
\nonumber\\
{\bf A}
&=&
{-}\!\left(2J_1{+}2J_{\sf x}{+}\frac{J_2}{2}\right)\!{\bf 1}{+}J_1{\bf K}
\ ,
\;
{\bf B}{=}\frac{J_2}{2}{\bf 1}{+}J_{\sf x}{\bf K}
\ ,
\nonumber\\
{\bf K}
&=&
\left(
\begin{array}{ccc}
0 & \frac{1+{\rm e}^{-{\rm i}k_1}}{2} & \frac{1+{\rm e}^{-{\rm i}k_2}}{2} \\
\frac{1+{\rm e}^{{\rm i}k_1}}{2} & 0 & \frac{1+{\rm e}^{{\rm i}(k_1-k_2)}}{2} \\
\frac{1+{\rm e}^{{\rm i}k_2}}{2} & \frac{1+{\rm e}^{-{\rm i}(k_1-k_2)}}{2} & 0
\end{array}
\right)\!.
\end{eqnarray}
The eigenvalues of matrix ${\bf K}$ (\ref{04}) are as follows:
$\varkappa_1=-1$,
$\varkappa_{2,3}=(1\mp\sqrt{3+2\gamma_{\bf k}})/2$,
where $\gamma_{\bf k}=\cos k_1 +\cos k_2 +\cos(k_1-k_2)$.
Now, we may use the formulas for determinants of block matrices \cite{Mirsky1955,Silvester2000}.
Since $\bf A$ and $\bf B$ commute,
the sought eigenvalues of $\bf H$ follow from the equations:
${\rm det}({\bf A}+{\bf B}-\Lambda^{(i)}_{\bf k} {\bf 1})=0$, $i=1,2,3$
and
${\rm det}({\bf A}-{\bf B}-\Lambda^{(i)}_{\bf k} {\bf 1})=0$, $i=4,5,6$.
As a result,
\begin{eqnarray}
\label{05}
\Lambda^{(1)}_{\bf k}&=&-3\left(J_1+J_{\sf x}\right),
\nonumber\\
\Lambda^{(2,3)}_{\bf k}&=&-\frac{J_1+J_{\sf x}}{2}
\left(3\pm\sqrt{3+2\gamma_{\bf k}}\right),
\nonumber\\
\Lambda^{(4)}_{\bf k}&=&-\left(3J_1+J_{\sf x}+J_2\right),
\nonumber\\
\Lambda^{(5,6)}_{\bf k}&=&{-}\left(\frac{3J_1{+}5J_{\sf x}}{2}{+}J_2\right)
\mp
\frac{J_1{-}J_{\sf x}}{2}\sqrt{3{+}2\gamma_{\bf k}}.
\end{eqnarray}
In the presence of a magnetic field one has to add $-h{\cal N}$ to the ferromagnetic-state energy and replace $\Lambda^{(i)}_{\bf k}$ by $\Lambda^{(i)}_{\bf k}+h$.
Note that $\Lambda^{(1,2,3)}_{\bf k}$ depend on $J_1+J_{\sf x}=2J$ only and do not depend on $J_2$.
For the fully frustrated case $J_1=J_{\sf x}=J$, there are four flat bands:
$\Lambda^{(1)}_{\bf k}=-6J$ and $\Lambda^{(4,5,6)}_{\bf k}=-(4J+J_2)$.
Furthermore, 
$\Lambda^{(2,3)}_{\bf k}=-J(3\pm\sqrt{3+2\gamma_{\bf k}})$ 
remain dispersive 
and $\Lambda^{(2)}_{{\bf k}=0}$ touches the flat band $\Lambda^{(1)}_{\bf k}$.
The one-magnon spectrum is illustrated in Fig.~\ref{f03}.
Evidently, Eq.~(\ref{05}) is in perfect correspondence with the numerical results for $S^z=17$ ($N=36$), see Appendix~\ref{aa}. 

By inspection, the flat-band states with the energy $\Lambda^{(1)}_{\bf k}$ have the form
\begin{eqnarray}
\label{06}
{\propto}
\!\left[
\left({\rm e}^{{\rm i}k_2}{-}{\rm e}^{{\rm i}k_1}\right) t^-_{{\bf k}1} 
{+}
\left(1{-}{\rm e}^{{\rm i}k_2}\right) t^-_{{\bf k}2}
{+}
\left({\rm e}^{{\rm i}k_1}{-}1\right) t^-_{{\bf k}3}
\right]
\vert\!\uparrow\!{\ldots}\!\uparrow\rangle 
\nonumber\\
\end{eqnarray}
with $t^-_{{\bf k}\alpha}{=}s^-_{{\bf k}\alpha a}{+}s^-_{{\bf k}\alpha b}$.
In the ${\bf m}$-space, these are the hard-hegaxon states \cite{Schulenburg2002,Derzhko2004,Zhitomirsky2004,derzhko2007,derzhko2015} associated with the two nearest hexagons from different layers.
Furthermore, the flat-band states with the energies $\Lambda^{(4,5,6)}_{\bf k}$ to be denoted further by $\epsilon_0={-}(4J{+}J_2)$ are the singlets localized on the $J_2$ bonds, 
\begin{eqnarray}
\label{07}
\frac{1}{\sqrt{2}}
\left(
\vert\uparrow_{{\bf m}\alpha a}\rangle\vert\downarrow_{{\bf m}\alpha b}\rangle
-
\vert\downarrow_{{\bf m}\alpha a}\rangle\vert\uparrow_{{\bf m}\alpha b}\rangle
\right),
\end{eqnarray}
in the environment of all other fully polarized sites.
In the present study, we are interested in the case $J_2>2J$ (the large-$J_2$ regime), when the localized singlets, i.e., the localized magnons from the flat bands $\Lambda^{(4,5,6)}_{\bf k}$, are the lowest-energy one-magnon states.
Importantly, the local nature of the one-magnon flat-band states paves the way to the construction of  many-magnon ground states in the subspaces with ${\cal N}/3\le S^z<{\cal N}-1$.

Another route to determine and characterize the eigenstates of the frustrated bilayer Heisenberg Hamiltonian is to introduce the total spin operator on the $J_2$ bonds, i.e., ${\bf t}_{{\bf m}\alpha}={\bf s}_{{\bf m}\alpha a}+{\bf s}_{{\bf m}\alpha b}$ \cite{Honecker2000}. Moreover, we also introduce another spin operator on the $J_2$ bonds: ${\bf d}_{{\bf m}\alpha}={\bf s}_{{\bf m}\alpha a}-{\bf s}_{{\bf m}\alpha b}$. One can easily convinced oneself that the Hamiltonian (\ref{01}) in terms of these operators becomes as follows:
\begin{eqnarray}
\label{08}
H
&=&
\sum_{{\bf m}\alpha}
\left[{-}ht^z_{{\bf m}\alpha}{+}\frac{J_2}{2}\left({\bf t}^2_{{\bf m}\alpha}{-}\frac{3}{2}\right)\right]
\\
&&{+}\sum_{\langle{\bf m}\alpha\,{\bf n}\beta\rangle}\!\left(\frac{J_1{+}J_{\sf x}}{2}{\bf t}_{{\bf m}\alpha}{\cdot}{\bf t}_{{\bf n}\beta}
{+}\frac{J_1{-}J_{\sf x}}{2}{\bf d}_{{\bf m}\alpha}{\cdot}{\bf d}_{{\bf n}\beta}\right)
\nonumber
\ .
\end{eqnarray}
For the fully frustrated case $J_1=J_{\sf x}=J$, the last term in Eq.~(\ref{08}) drops out and the Hamiltonian depends only on the total spin operators on the $J_2$ bonds ${\bf t}_{{\bf m}\alpha}$.
Since ${\bf t}^2_{{\bf m}\alpha}$ commutes with the Hamiltonian for all ${\bf m}\alpha$ (local integrals of motion), the Hamiltonian eigenstates can be labeled by the set of ${\cal N}$ good quantum numbers ${\bf t}^2=t(t+1)$ assigned to each $J_2$ bond. 
Obviously, $t$ may acquire only two values, $t=0$ (singlet) or $t=1$ (triplet).
In the latter case, when $t=1$ for all ${\bf m}\alpha$, one arrives at the $S=1$ kagome-lattice Heisenberg antiferromagnet, which is among the reference models of frustrated quantum magnetism. This model represent the low-energy physics of the $S=1/2$ kagome-lattice bilayer for the ferromagnetic interlayer interaction $J_2<0$.
In the present study, however, we focus on the case of large enough antiferromagnetic interlayer interaction $J_2>2J>0$ and $h\geq 0$. Then localized singlets and (polarized) triplets dominate the low-temperature properties and one arrives at classical statistical mechanics problems.   

Consider, for example, the eigenstates with only one $t=0$ and all others are polarized triplets, cf. Eq.~(\ref{07}). Such eigenstates correspond to the flat-band states with the energy $\epsilon_0=-(4J+J_2)$. Indeed, each such a state has the energy $E_{\rm FM}-4J-J_2$, where  $E_{\rm FM}=(2J+J_2/4){\cal N}$ is the energy of the ferromagnetic state (polarized triplets at all $J_2$ bonds), cf. the first term in Eq.~(\ref{03}).

\begin{figure}
\includegraphics[width=\columnwidth]{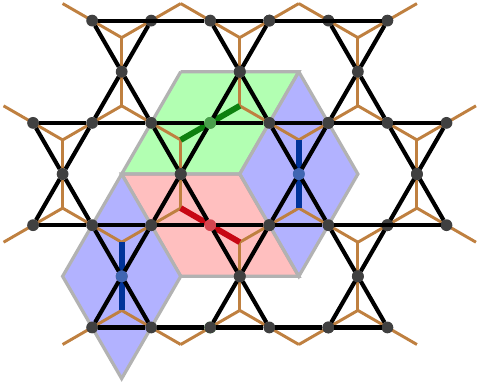}\\
\vspace{1mm} 
\includegraphics[width=\columnwidth]{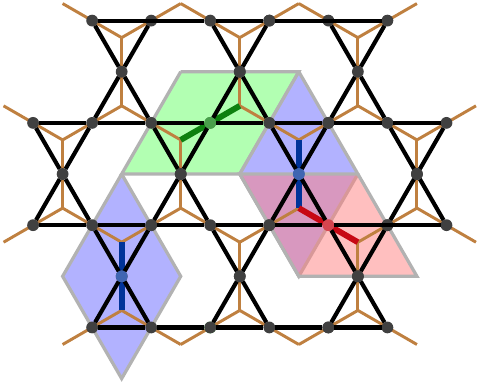} 
\caption{Top: Pictorial representation of many-magnon ground states for $S^z={\cal N}{-}1,\ldots,2{\cal N}/3$. Localized singlets (denoted by colored discs) obey hard-rhombi rule on the underlying kagome lattice (thick black) and therefore can be presented as hard-core rhombi. They also can be viewed as hard-core dimers on an auxiliary honeycomb lattice (thin brown).
Bottom: Partially overlapping (soft) rhombi represent other localized eigenstates of the fully frustrated Hamiltonian (\ref{08}).}
\label{f04}
\end{figure}

Because of the local character of the singlet state, there is a class of many-magnon states, which are located on $J_2$ bonds sufficiently far from each other. More precisely, this class of states contain $n=2,\ldots,n_{\max}$, $n_{\max}={\cal N}/3$ $J_2$ bonds with $t=0$ (the rest ${\cal N}-n$ $J_2$ bonds carry polarized triplets) and satisfy the geometrical restriction: The singlets are not neighbors (hard-core rule), see Fig.~\ref{f04}, top. Alternatively, such many-magnon states can be viewed as hard-core dimers on an auxiliary  honeycomb lattice, see Fig.~\ref{f04}, top. The energy of hard-rhombi states is $E_{\rm FM}+n\epsilon_0$. And such states are the ground states in the subspaces with $S^z={\cal N}-1,\ldots,{\cal N}-n_{\max}$, $n_{\max}={\cal N}/3$ if $J_2>2J$. This picture perfectly agrees with numerical data for $S^z=17,\ldots,12$ ($N=36$), see Appendix~\ref{aa}. 

There are two more classes of localized states which are eigenstates of the fully frustrated Hamiltonian (\ref{08}). That is, i) the states with $n=2,\ldots,2n_{\max}$ singlets on the $J_2$ bonds forming patterns with one (or more) pair(s) which are neighbors, see Fig.~\ref{f04}, bottom, and ii) the states with $n=3,\ldots,3n_{\max}$ singlets on the $J_2$ bonds forming patterns with one (or more) triple(s) of singlets which are neighbors. These states can visualized as containing partially overlapping two rhombi (see Fig.~\ref{f04}, bottom) or partially overlapping three rhombi. The energy penalty for each overlap is $J$.

For sufficiently large $J_2/J$ the three classes of localized eigenstates described above (their total number is $2^{\cal N}$) are the ground states and low-lying excited states of the model at hand. Again, this picture of overlapping rhombi is confirmed by numerical data for $N=36$, see Appendix~\ref{aa}.

\subsection{Mapping onto classical lattice gases}
\label{s3b}

\begin{table}
\caption{\label{t01}
Counting the number of $n$ hard-rhombi spatial configurations on periodic kagome lattices of ${\cal N}$ sites. For ${\cal N}=18$ two clusters are determined by the edge vectors ${\bf a}_1=(6,0)$, ${\bf a}_2=(2,2\sqrt{3})$ (${\cal N}=18_a$) and ${\bf a}_1=(6,0)$, ${\bf a}_2=(0,2\sqrt{3})$ (${\cal N}=18_b$). For ${\cal N}=48$ two clusters are determined by the edge vectors ${\bf a}_1=m{\bf e}_1$, ${\bf a}_2=n{\bf e}_2$, $m=n=4$ (${\cal N}=48_a$) and ${\bf a}_1=m{\bf e}_1$, ${\bf a}_2=-{\bf e}_1+n{\bf e}_2$, $m=n=4$ (${\cal N}=48_b$).}
\begin{ruledtabular}
\vspace{1mm}
${\cal{N}}=18$
\vspace{1mm}
\begin{tabular}{r||rr} 
~$n$~ & $g_{18_a}(n)$, ~${\cal N}=18_a$~ & $g_{18_b}(n)$, ~${\cal N}=18_b$~ \\
\hline \hline
~1~ &  ~18~ &  ~18~ \\ 
~2~ & ~117~ & ~117~ \\
~3~ & ~336~ & ~336~ \\
~4~ & ~417~ & ~420~ \\
~5~ & ~186~ & ~192~ \\
~6~ &  ~17~ &  ~20~ \\
\end{tabular}
\vspace{3mm}
${\cal{N}}=48$
\vspace{1mm}
\begin{tabular}{r||rr} 
~$n$~ & $g_{48_a}(n)$, ~${\cal N}=48_a$~ & $g_{48_b}(n)$, ~${\cal N}=48_b$~ \\
\hline \hline
~1~ & ~48~ & ~48~ \\ 
~2~ & ~1~032~ & ~1~032~ \\
~3~ & ~13~136~ & ~13~136~ \\
~4~ & ~110~244~ & ~110~244~ \\
~5~ & ~643~056~ & ~643~056~ \\
~6~ & ~2~677~896~ & ~2~677~896~ \\
~7~ & ~8~052~432~ & ~8~052~432~ \\
~8~ & ~17~486~550~ & ~17~486~558~ \\
~9~ & ~27~158~096~ & ~27~158~224~ \\
~10~ & ~29~567~736~ & ~29~568~536~ \\
~11~ & ~21~843~888~ & ~21~846~384~ \\
~12~ & ~10~417~700~ & ~10~421~844~ \\
~13~ & ~2~969~616~ & ~2~973~264~ \\
~14~ & ~446~136~ & ~447~704~ \\
~15~ & ~27~952~ & ~28~208~ \\
~16~ & ~417~ & ~417~ \\
\end{tabular}
\end{ruledtabular}
\end{table}

The hard-rhombi picture is a good starting point to count the eigenstates described above. In Table~\ref{t01} we report the number of spatial configurations of $n$ hard rhombi on the periodic ${\cal N}$-site kagome lattice of two shapes, $g_{{\cal N}_a}(n)$ and $g_{{\cal N}_b}(n)$. These numbers match perfectly the ground-state degeneracy ${\cal W}_{\rm GS}(S^z)$ for the initial frustrated quantum spin system (\ref{01}) of $N=36_a$ sites and $N=36_b$ sites (see Appendix~\ref{aa}). However, for hard rhombi we can examine larger systems, see the results for ${\cal N}=48$ in Table~\ref{t01}, whereas the corresponding initial frustrated quantum spin system of $N=96$ sites is far beyond reachable sizes by numerics. More results on hard rhombi are collected in Appendix~\ref{ab}.

Hard rhombi on the kagome lattice can be treated similarly to other lattice gases of hard-core objects \cite{Baxter1982,Frenkel1992,Lafuente2004,Fernandes2007}.
More specifically, one can introduce the grand canonical partition function
\begin{eqnarray}
\label{09}
\Xi(z,{\cal N})=\sum_{n=0}^{n_{\max}}z^n g_{\cal N}(n),
\end{eqnarray}
where $z={\rm e}^{\mu/T}$ is the activity and $g_{\cal N}(n)$ is the canonical partition function of $n$ hard rhombi on the kagome lattice of ${\cal N}$ sites or, in other words, the number of allowed spatial configurations. 

\begin{figure}
\includegraphics[width=\columnwidth]{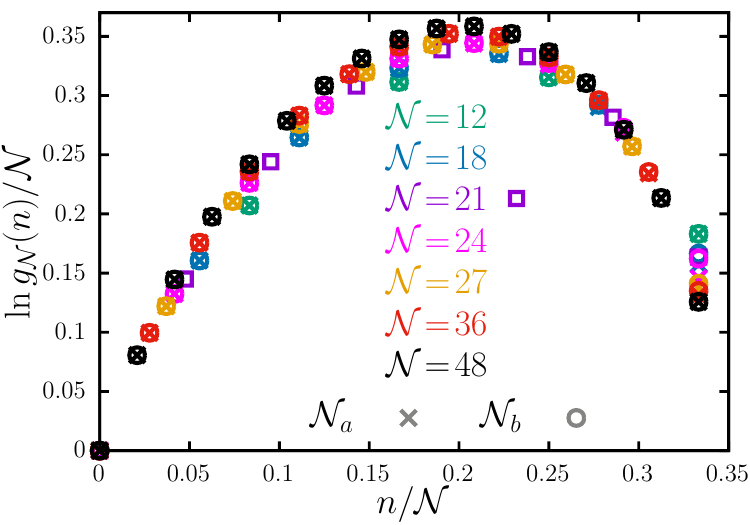}\\
\vspace{1mm} 
\includegraphics[width=\columnwidth]{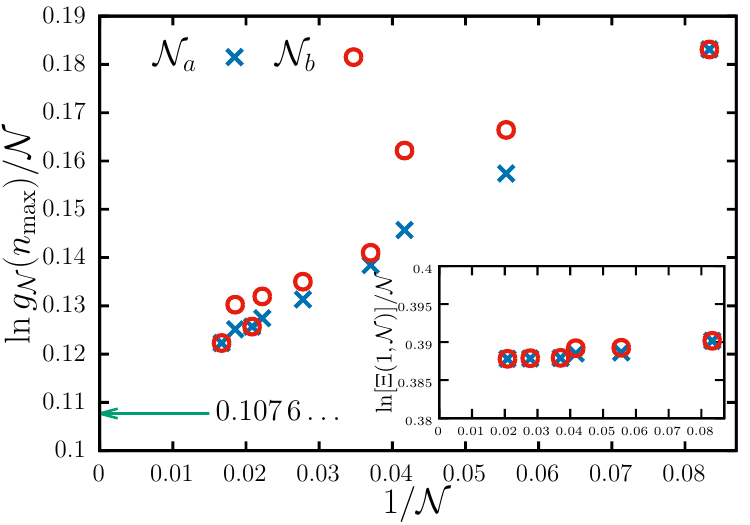} 
\caption{Hard-rhombi description of the kagome-lattice bilayer. ${\cal N}$ denotes the number of kagome-lattice sites; $n$ denotes the number of hard rhombi.
Top: $\ln g_{\cal N}(n)/{\cal N}$ versus $n/{\cal N}$ for various periodic clusters. 
Bottom: $s_0({\cal N}/3,{\cal N})$ (\ref{10}) versus $1/{\cal N}$; 
we get in the thermodynamic limit, linear fit, $0.106\,9\ldots$ (${\cal N}_a$) and $0.110\,6\ldots$ (${\cal N}_b$). 
Inset in the bottom panel: $\ln\kappa(1)$ (\ref{11}) versus $1/{\cal N}$; 
we get in the thermodynamic limit, linear fit, $0.386\,7\ldots$ (${\cal N}_a$) and $0.386\,9\ldots$ (${\cal N}_b$).} 
\label{f05}
\end{figure}

Clearly, $g_{\cal N}(n)$ which is related to the ground-state degeneracy yields the residual ground-state entropy through the relation  
\begin{eqnarray}
\label{10}
s_0(n,{\cal N})=\frac{\ln g_{\cal N}(n)}{{\cal N}}.
\end{eqnarray}
In the upper panel of Fig.~\ref{f05} we report the dependence of the residual ground-state entropy $s_0$ on the density of hard rhombi on the kagome lattice $n/{\cal N}$.

\begin{table}
\caption{\label{t02}
Numbers of spatial configurations of $n_{\max}={\cal N}/3$ hard rhombi on the kagome lattice of ${\cal N}$ sites for two clusters with edge vectors ${\bf a}_1=m{\bf e}_1$, ${\bf a}_2=n{\bf e}_2$, $mn={\cal N}/3$ (${\cal N}_a$) and ${\bf a}_1=m{\bf e}_1$, ${\bf a}_2=-{\bf e}_1+n{\bf e}_2$, $mn={\cal N}/3$ (${\cal N}_b$).}
\begin{ruledtabular}
\begin{tabular}{r||rr}
~$n_{\text{max}}$~ & $g_{{\cal N}_a}(n_{\max})$, ~${\cal N}_a$~ & $g_{{\cal N}_b}(n_{\max})$, ~${\cal N}_b$~ \\
\hline \hline
~${\cal{N}}=12$~ &     ~9~ &     ~9~ \\ 
~${\cal{N}}=18$~ &    ~17~ &    ~20~ \\
~${\cal{N}}=24$~ &    ~33~ &    ~49~ \\
~${\cal{N}}=27$~ &    ~42~ &    ~45~ \\
~${\cal{N}}=36$~ &   ~113~ &   ~129~ \\
~${\cal{N}}=45$~ &   ~309~ &   ~379~ \\
~${\cal{N}}=48$~ &   ~417~ &   ~417~ \\
~${\cal{N}}=54$~ &   ~860~ & ~1~133~ \\
~${\cal{N}}=60$~ & ~1~537~ & ~1~537~ \\
\end{tabular}
\end{ruledtabular}
\end{table}

The case of $n=n_{\max}$ (full covering) deserves more discussion.
In Table~\ref{t02} we report the results for $g_{\cal N}(n_{\max})$, $n_{\max}={\cal N}/3$ for periodic lattices up to ${\cal N}=60$ and in the lower panel of Fig.~\ref{f05} we use these data to plot the residual ground-state entropy $s_0(n_{\max},{\cal N})=[\ln g_{\cal N}(n_{\max})]/{\cal N}$ (i.e., for $n/{\cal N}=n_{\max}/{\cal N}=1/3$) as a function of $1/{\cal N}$ to illustrate what happens in the limit ${\cal N}\to \infty$. The simplest linear fit results in, depending on the specific boundary conditions, $0.106\,9\ldots$ or $0.110\,6\ldots$. Both numbers for the estimate of $s_0({\cal N}/3,{\cal N})$ in the thermodynamic limit are close to $0.107\,6\ldots$, and this issue is discussed in the next paragraph.

From another perspective, as is evident from Fig.~\ref{f04}, the kagome-lattice hard-rhombi problem can be mapped onto the honeycomb-lattice hard-dimer problem with the relation ${\sf N}=2{\cal N}/3$ between the numbers of honeycomb-lattice sites ${\sf N}$ and kagome-lattice sites ${\cal N}$ and the number of dimers $0,1,\ldots, {\sf N}/2$ corresponds to the number of rhombi $0, 1,\ldots,{\cal N}/3$. 
Moreover, it is known that the number of dimer coverings (close-packed dimers) ${\sf W}$ on the infinite periodic honeycomb lattice with ${\sf N}\to\infty$ sites is $\ln {\sf W}/({\sf N}/2)=0.323\,0\ldots$ \cite{Wu2006}.
Since ${\sf W}=g_{\cal N}(n_{\max})$, the limiting value of $s_0(n_{\max},{\cal N})$ is three times smaller than $0.323\,0\ldots$ in agreement with the lower panel of Fig.~\ref{f05}.

Another function $\kappa(z)$, which has been intensively investigated for lattice gases of hard-core objects, is defined as
\begin{eqnarray}
\label{11}
\ln\kappa(z)=\frac{\ln\Xi(z,{\cal N})}{{\cal N}}.
\end{eqnarray}
Its values at $z=1$, i.e., $\ln\kappa(1)=\ln[\sum_{n=0}^{n_{\max}}g_{\cal N}(n)]/{\cal N}$ were estimated for several models in the past.
Namely,
for hard squares on the square lattice
$\ln\kappa(1)=0.407\,4\ldots$,
for hard hexagons on the honeycomb lattice
$\ln\kappa(1)=0.435\,9\ldots$,
for hard hexagons on the triangular lattice
$\ln\kappa(1)=0.333\,2\ldots$,
see, e.g., Ref.~\cite{Baxter1999}.
As can be seen from the inset in the lower panel of Fig.~\ref{f05},
for hard rhombi on the kagome lattice we obtain $\ln\kappa(1)\approx 0.387$.
As a result, we conclude that within the hard-rhombi picture we take into account $\kappa(1)^{\cal N}\approx{\rm e}^{0.387{\cal N}}$ states, whereas the soft-rhombi picture accounts for $2^{{\cal N}}\approx{\rm e}^{0.693{\cal N}}$ states. Although these localized states constitute only a part of the total $2^{2{\cal N}}\approx{\rm e}^{1.386{\cal N}}$ states for the initial model (\ref{01}), they dominate the low-temperature magnetothermodynamics if $J_2>2J$.

\subsection{Low-temperature magnetothermodynamics}
\label{s3c}

After counting the hard-rhombi states, we can go one step further and elaborate their contribution to observable properties.  
In the presence of an external magnetic field $h>0$, the energy of the $n$ hard-rhombi state is $E_n(h)=E_{\rm FM}-{\cal N}h+n(\epsilon_0+h)$. The number of hard-rhombi states $g_{\cal N}(n)$ is equal to the number of allowed spatial configurations of $n$ hard rhombi on the ${\cal N}$-site kagome lattice, that is, to their canonical partition function. As a result, the contribution of localized singlets satisfying the hard-core rule to the partition function of the frustrated quantum spin system $Z_{\rm lm}(T,h,N)$ is related to the grand canonical partition function of hard-core rhombi on kagome lattice $\Xi(z,{\cal N})$ (\ref{09})
\begin{eqnarray}
\label{12}
Z_{\rm lm}(T,h,N)=\sum_{n=0}^{n_{\rm max}}g_{\cal N}(n){\rm e}^{-\frac{E_n(h)}{T}}
\nonumber\\
=
{\rm e}^{-\frac{E_{\rm FM}-{\cal N}h}{T}}\sum_{n=0}^{n_{\rm max}}z^ng_{\cal N}(n)
=
{\rm e}^{-\frac{E_{\rm FM}-{\cal N}h}{T}}\Xi(z,{\cal N}),
\nonumber\\
z={\rm e}^{\frac{\mu}{T}},
\;\;\;
\mu=h_{\rm sat}-h,
\;\;\;
h_{\rm sat}=-\epsilon_0=4J+J_2.
\end{eqnarray} 
This is a dominant contribution at low temperatures and high fields.  

It is convenient to represent the grand canonical partition function of hard-core rhombi on a kagome lattice $\Xi(z,{\cal N})$, which enters Eq.~(\ref{12}), in terms of the on-site occupation numbers $n_i=0,1$:
\begin{eqnarray}
\label{13}
\Xi(z,{\cal N}){=}
\sum_{n_1=0,1}\!\ldots\!\sum_{n_{\cal N}=0,1}
z^{n_1+\!\ldots\!+n_{\cal N}}{\rm e}^{-\frac{V\sum_{\langle ij \rangle}n_in_j}{T}}
\end{eqnarray}
(the sum in the exponent runs over all nearest neighbors on the kagome lattice).
Sending here $V\to+\infty$ (infinite repulsion), we restrict the allowed configurations to the hard-rhombi patterns. 
Finite repulsion $V$ accounts for overlapping rhombi states and setting $V=J$ we reproduce correctly the energy of these states.
Clearly, in Eq.~(\ref{13}) we face the grand canonical partition function of a classical lattice-gas model on the kagome lattice defined by the Hamiltonian
\begin{eqnarray}
\label{14}
{\cal H}(\{n_i\})=-\mu\sum_{i} n_i +V\sum_{\langle ij \rangle}n_in_j,
\nonumber\\
\mu=h_{\rm sat}-h, 
\;
h_{\rm sat}=4J+J_2
\end{eqnarray} 
with infinite nearest-neighbor repulsion $V\to\infty$ (hard-core rhombi) or $V=J$ (partially overlapping or soft-core rhombi). 
The lattice-gas model of rhombi (\ref{14}) can be cast into the antiferromagnetic Ising model in a field after the change $n_i=1/2-T_i^z$:
\begin{eqnarray}
\label{15}
\!\!\!
{\cal H}(\{T^z_i\}){=}\frac{V{-}\mu}{2}{\cal N}
{+}\left(\mu{-}2V\right)\sum_{i} T^z_i {+}V\sum_{\langle ij \rangle}T^z_iT^z_j.
\end{eqnarray}
Kagome-lattice Ising models were examined already 
more than seventy years ago \cite{Syozi1951,Kano1953}. 

It is worth stressing that the lattice-gas model (and the Ising model)  possesses a symmetry with respect to the change of variables $n_i$ to $\overline{n}_i=1-n_i$  (equivalently $T_i^z$ to $-T_i^z$): The thermodynamic quantities for $h_{\rm sat}-h$ are straightforwardly related to those for $h-h_1$.
Another symmetry occurs at $h_2=J_2+2J_1$: Again, the thermodynamic quantities for $h_{2}+\Delta h$ are related to those for $h_2-\Delta h$.

Now, we use the classical lattice models to corroborate the suggested description of low-lying Hamiltonian eigenstates by comparison with numerics. More precisely, we perform direct calculations for the lattice models given in Eq.~(\ref{14}) or Eq.~(\ref{15}) with $V\to\infty$ and $V=J$, to obtain various quantities and compare them to the finite-lattice results for the initial quantum spin model (\ref{01}). In particular, we consider the magnetization and the susceptibility per site of the initial model (\ref{01}) 
\begin{eqnarray}
\label{16}
m(T,h)=\frac{1}{2}+\frac{T}{2\cal N}\frac{\partial \ln \Xi(T,\mu,{\cal N})}{\partial h},
\nonumber\\
\chi(T,h)=\frac{\partial m(T,h)}{\partial h}
\end{eqnarray}
along with the entropy and the specific heat per site of the initial model (\ref{01}) 
\begin{eqnarray}
\label{17}
s(T,h)=\frac{\ln \Xi(T,\mu,{\cal N})}{2{\cal N}}+\frac{T}{2{\cal N}}\frac{\partial \ln \Xi(T,\mu,{\cal N})}{\partial T},
\nonumber\\
c(T,h)=T\frac{\partial s(T,h)}{\partial T},
\end{eqnarray}
where $\Xi(T,\mu,{\cal N})$ is the grand canonical partition function of a lattice gas.

\begin{figure*}
\includegraphics[width=0.9\columnwidth]{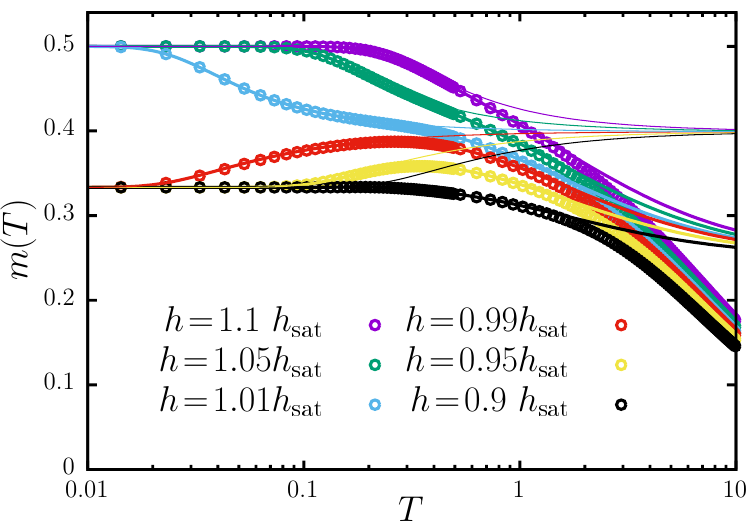} 
\includegraphics[width=0.9\columnwidth]{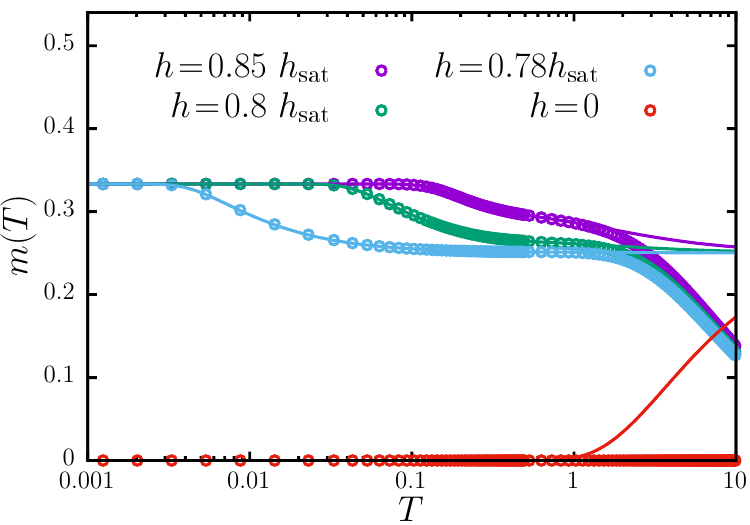} 
\includegraphics[width=0.9\columnwidth]{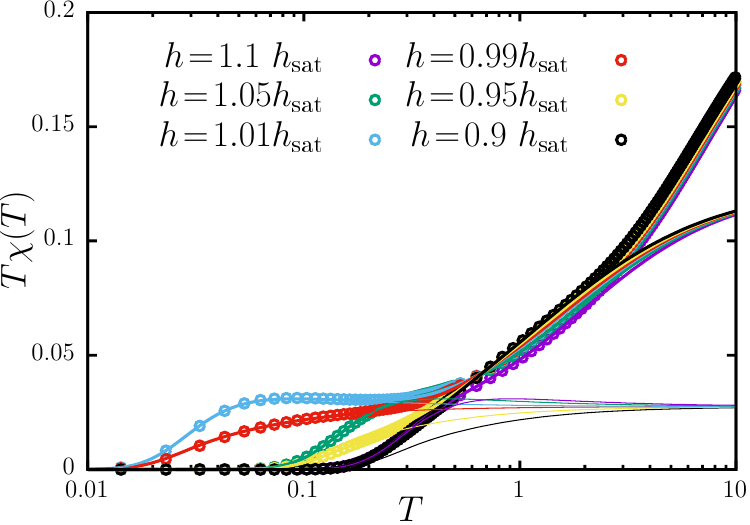} 
\includegraphics[width=0.9\columnwidth]{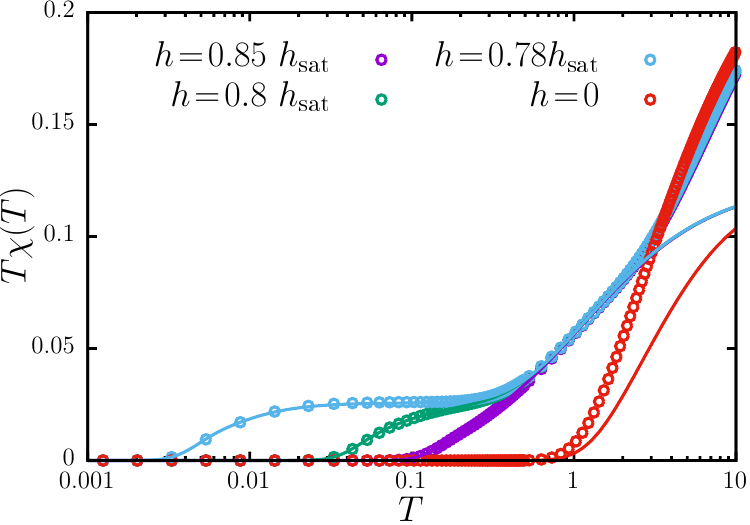} 
\includegraphics[width=0.9\columnwidth]{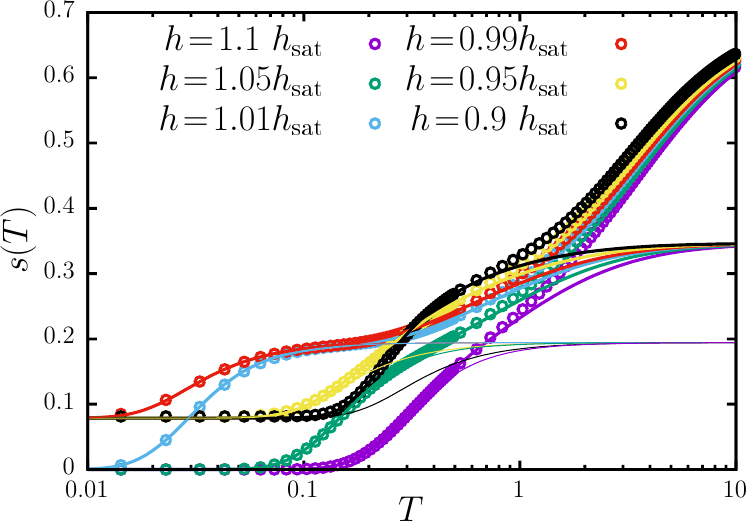} 
\includegraphics[width=0.9\columnwidth]{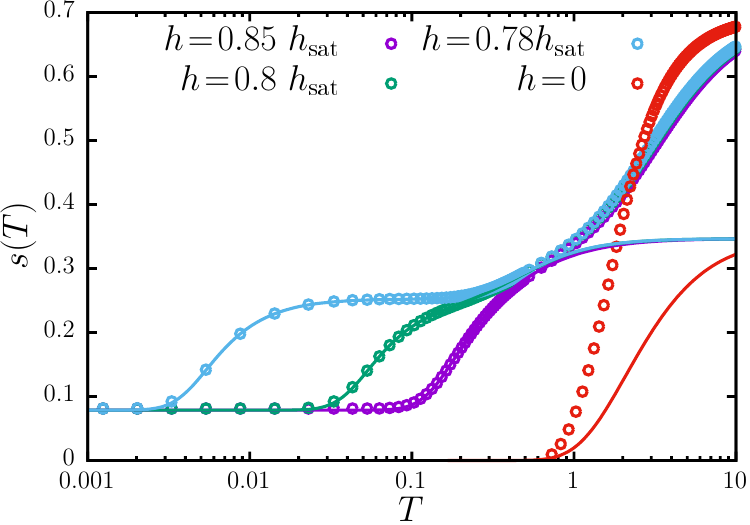} 
\includegraphics[width=0.9\columnwidth]{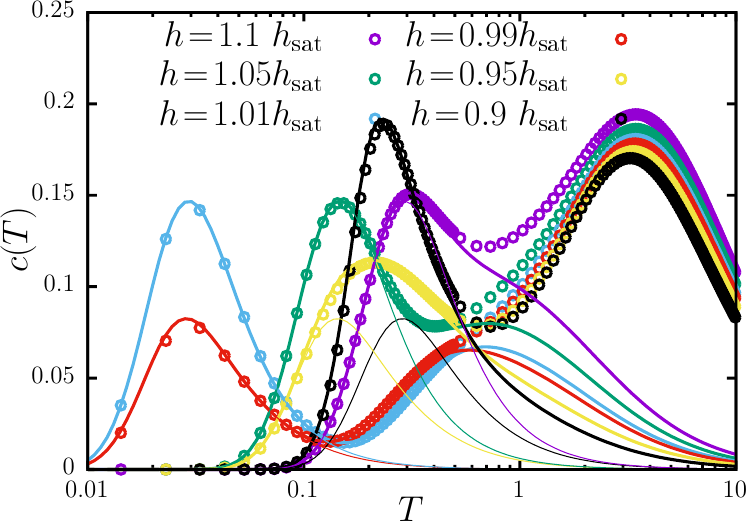} 
\includegraphics[width=0.9\columnwidth]{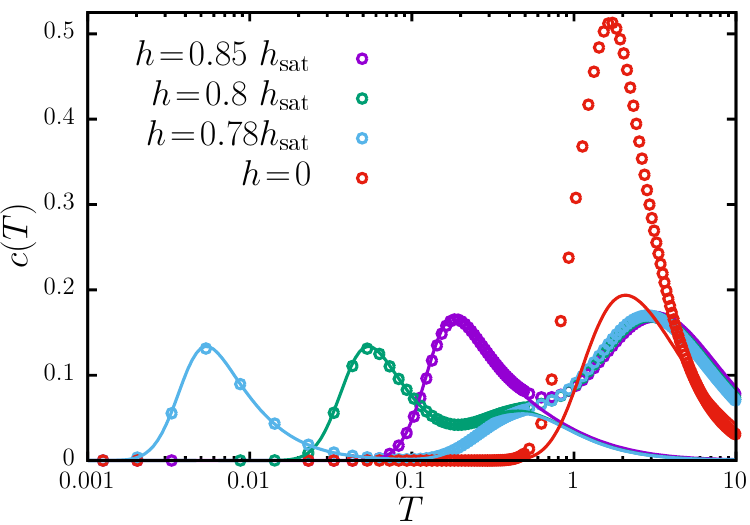} 
\caption{Hard- and soft-core rhombi predictions (thin and thick curves) against finite-temperature Lanczos method data (circles) for thermodynamic characteristics, $J=1$, $J_2=5$, $N=36_b$. 
From top to bottom: Magnetization $m$, susceptibility $\chi$, entropy $s$, and specific heat $c$ per site of the initial model (\ref{01}). 
Left column: High fields $h=9.9$, $9.45$, $9.09$, $8.91$, $8.55$, $8.1$. 
Right column: Moderate fields $h=7.65,7.2,7.02$ and $h=0$.
Hard-core rhombi work up to $T\approx0.15$; soft-core rhombi work up to $T\approx0.45$.}
\label{f06}
\end{figure*}

Let us begin with the temperature dependencies around saturation, see the left column in Fig.~\ref{f06}.
According to numerics for $N=36_b$ (circles), $m(T)$ for $h>h_{\rm sat}$ decreases as $T$ grows, while $m(T)$ for $h\lessapprox h_{\rm sat}$ slightly increases as $T$ deviates from zero and then decreases as $T$ grows further. This is in accordance with the expected smearing out of the ground-state magnetization jump at the saturation field $h_{\rm sat}$. The hard-rhombi description (thin curves) works perfectly well in the low-temperature region but fails above $T\approx0.2$ and cannot reproduce vanishing magnetization to zero in the high-temperature limit. Instead, as is obvious from Eq.~(\ref{16}), the hard-rhombi prediction in the high-temperature limit tends to a nonzero value, which is related to the averaged density of hard rhombi $\langle n\rangle/{\cal N}$ at $z{=}1$ (according to Table~\ref{t01}, $(\langle n\rangle/{\cal N})\vert_{z=1}\approx 0.202$ for ${\cal N}=18$). Soft rhombi (thick curves) provide a correct description up to $T\approx1$ but for higher temperatures tends again to a nonzero value, which is related to the averaged density of soft rhombi $\langle n\rangle/{\cal N}$ at $T{\to}\infty$, $(\langle n\rangle/{\cal N})\vert_{T\to\infty}{=}1/2$. 
The next panel in the left column of Fig.~\ref{f06} concerns the temperature dependence of the uniform susceptibility. In the high-temperature limit it obeys the Curie law: $\chi(T{\to}\infty){\to}1/(4T)$; in the temperature range $T<10$ this regime is not achieved yet (circles). Hard- and soft-rhombi correctly reproduce the low-temperature behavior as well as the $1/T$ dependence at high temperatures, however, with a constant smaller than $1/4$.
Furthermore, the temperature profiles of entropy have clear interpretation. For $h>h_{\rm sat}$, $s(T{=}0){=}0$, but for $h_{2}<h<h_{\rm sat}$, $s(T{=}0){=}\ln {\cal W}_{\rm GS}/N$, ${\cal W}_{\rm GS}=g_{\cal N}({\cal N}/3)$, i.e., is $\approx0.083$ for $N=36_b$. Rhombi pictures perfectly reproduce the low-temperature region. On the other hand, as $T\to\infty$, the entropy tends to $\ln 2$ (exact value), but to $\ln 2/2$ (soft rhombi) or $\ln\kappa(1)/2$ (hard rhombi).
We finish the discussion with a comparison of the temperature dependence of the specific heat obtained for the initial and effective models, $N=36_b$ and ${\cal N}=18_b$, see the lowest panel in the left column in Fig.~\ref{f06}. The lower peak of $c(T)$, the position of which is controlled by the energy scale $\vert h_{\rm sat}-h\vert$, perfectly matches the rhombi predictions. The main peak of $c(T)$ at $T\approx 4$ is beyond capabilities of the rhombi description. 

The temperature dependencies of $m$, $\chi$, $s$, and $c$ for lower fields are shown in  the right column of Fig.~\ref{f06}. The reported comparisons demonstrate that the soft-rhombi description works even at $h=0$ as long as the temperature does not exceed $\approx 0.45$; the hard-rhombi picture (not shown here) in general fails further away from $h_{\rm sat}$. 

Overall, the comparison between hard/soft-rhombi predictions and exact numerics reported in Fig.~\ref{f06} indicates the region of validity of the effective-model description which obviously concerns sufficiently low temperatures only. In this region of validity, such a description opens new possibilities to examine the frustrated quantum spin model (\ref{01}) by means of the classical statistical mechanics. Besides direct calculations for ${\cal N}$ about 60 (see Appendix~\ref{ab}), classical Monte Carlo simulations for much larger ${\cal N}$ can be used, too.

\subsection{Do the bulk properties depend on the form?}
\label{s3d}

We conclude this section with an interesting consequence of the established dimer representation for the low-energy physics of the frustrated quantum spin model (\ref{01}). Usually it is implied that the bulk properties are insensitive to the boundary conditions in the limit of a large system. Remarkably, in 1961 P.~Kasteleyn \cite{Kasteleyn1961}, while studying dimer arrangements on a square lattice, expressed doubts on the independence of the bulk free energy on boundary conditions (for more recent publications see Refs.~\cite{Korepin2000,Cohn2001}). More relevant in the present context is the paper by V.~Elser \cite{Elser1984}, who considered the dimer problem on the honeycomb lattice with boundary and demonstrated explicitly that the bulk entropy per dimer in the thermodynamic limit depends on the shape of the boundary.

\begin{figure}
\includegraphics[width=\columnwidth]{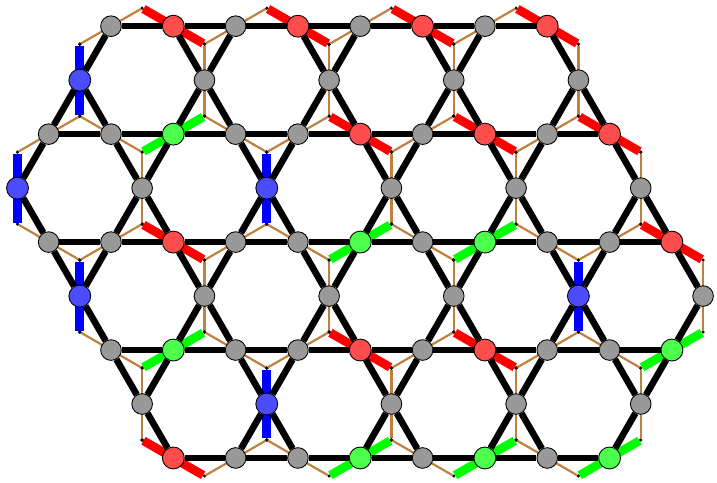}\\ 
\vspace{3mm}
\includegraphics[width=\columnwidth]{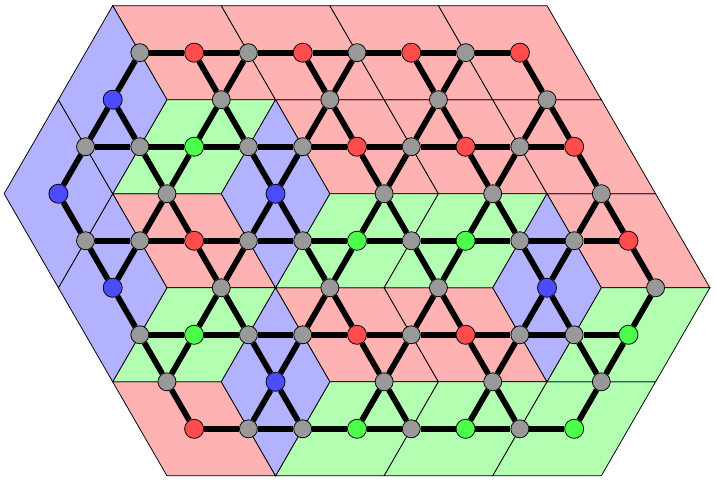} 
\caption{General hexagon with the sides $k=4$, $l=3$, and $m=2$ considered for calculating the bulk entropy for the dimer problem on the honeycomb lattice by Elser \cite{Elser1984} (top) and the corresponding rhombi on the kagome lattice (bottom).}
\label{f07}
\end{figure}

To be precise, for periodic (toroidal) boundary conditions the bulk entropy per dimer $\ln {\sf W}/({\sf N}/2)$ has been known since the 1950s, and it equals $0.323\,0\ldots$ \cite{Wu2006}.
Elser studied the dimer problem on a general hexagon defined by the sides $k$, $l$, and $m$, see Fig.~\ref{f07}. In particular, using MacMahon's results for the combinatorial problem of `plane partitions', he provides the limiting value of the entropy per dimer $\ln {\sf W}/({\sf N}/2)=S_{klm}/({\sf N}/2)$ as ${\sf N}\to\infty$:
\begin{eqnarray}
\label{18}
\frac{S_{klm}}{\frac{{\sf N}}{2}}\sim s(x,y,z)
\ ,\;\;\;\;\;
\\
\frac{{\sf N}}{2}=kl+lm+mk
\ ,
\nonumber\\
s(x,y,z){=}\frac{1}{2(xy{+}yz{+}zx)}
\left[
x^2\ln{x}
{-}(1{-}x)^2\ln{(1{-}x)}
\right.
\nonumber\\
\left.
{+}y^2\ln{y}
{-}(1{-}y)^2\ln{(1{-}y)}
{+}z^2\ln{z}
{-}(1{-}z)^2\ln{(1{-}z)}
\right],
\nonumber\\
x=\frac{k}{n},
\;\;\;
y=\frac{l}{n},
\;\;\;
z=\frac{m}{n},
\;\;\;
n=k+l+m.
\nonumber
\end{eqnarray}
This formula demonstrates how the bulk entropy per dimer depends on the shape of the boundary.
Moreover, the maximum of the entropy per dimer $s(x,y,z)$ is obtained when the boundary is a regular hexagon, i.e., $s(1/3,1/3,1/3)=(3/2)\ln 3 - 2 \ln 2\approx0.262$ which is smaller than the value for periodic boundary conditions. And the entropy per dimer goes to zero if, e.g., $x=1$, $y=z=0$.

Since the dimer problem on the honeycomb lattice is equivalent to the hard-rhombi covering of the kagome lattice and the residual ground-state entropy for the 2/3 plateau state $s_0({\cal N}/3,{\cal N})$ (\ref{10}) is three times smaller than $s(x,y,z)$, see above, we conclude that for $h_2<h<h_{\rm sat}$ the residual ground-state entropy (\ref{17}) $s(T=0,h\lessapprox h_{\rm sat})=s(x,y,z)/6$ depends on the shape of thermodynamically large open-boundary (hexagonal) system, see Eq.~(\ref{18}).  
On the other hand, $s(T)=\ln 2-\int_{T}^{\infty}{\rm d}{\tt T}c({\tt T})/{\tt T}$ and, as a result, we have the following sum rule:
\begin{eqnarray}
\label{19}
\int_{0}^{\infty}{\rm d}{\tt T}\frac{c({\tt T})}{{\tt T}}=\ln 2-s(T=0).
\end{eqnarray}
Since $s(T{=}0)$ on the r.h.s.\ of Eq.~(\ref{19}) depends on the shape of the thermodynamically large open-boundary (hexagonal) system, $c(T)$ in the l.h.s. of Eq.~(\ref{19}) should also depend on the shape.

\begin{figure}
\includegraphics[width=\columnwidth]{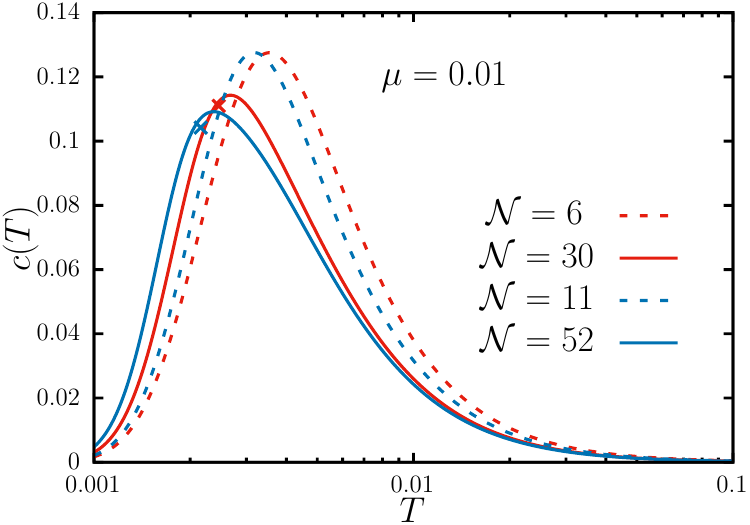} 
\caption{Temperature dependence of the specific heat ($\mu=h_{\rm sat}-h=0.01$) for two open hexagonal systems with $k:l:m=1:1:1$ (red) and $k:l:m=2:1:1$ (blue). $c(T)$ for ${\cal N}=6,11$ (dashed curves) and for ${\cal N}=30,52$ (solid curves). The position of the peak is linearly extrapolated for $1/{\cal N}\to0$ (crosses).}
\label{f08}
\end{figure}

It is not easy to check a dependence of $c(T)$ on the form of the thermodynamically large system having access only to rather small systems. In particular, the finite-temperature Lanczos method deals with $N=36,42$ (i.e., ${\cal N}=18,21$) only, while direct calculations for hard rhombi are restricted to ${\cal N}$ about $60$. 
In Fig.~\ref{f08} we report preliminary results for two hexagon shapes: $1:1:1$ (red curves) and $2:1:1$ (blue curves) as they follow from direct calculations for hard-rhombi systems with $\mu=h_{\rm sat}-h=0.01$ of sizes ${\cal N}=6,30$ (red dashed and solid curves) and ${\cal N}=11,52$ (blue dashed and solid curves). Moreover, we extrapolate the peak of $c(T)$ as $1/{\cal N}\to 0$ (linear extrapolation), which yields the red and blue crosses in Fig.~\ref{f08}. 
Our findings agree with the expectation about a shape-dependent $c(T)$. However, extensive numerical studies for larger systems are required to better understand this issue.

\section{Beyond the fully frustrated limit: Numerics versus effective theory}
\label{s4}

\begin{figure}
\includegraphics[width=\columnwidth]{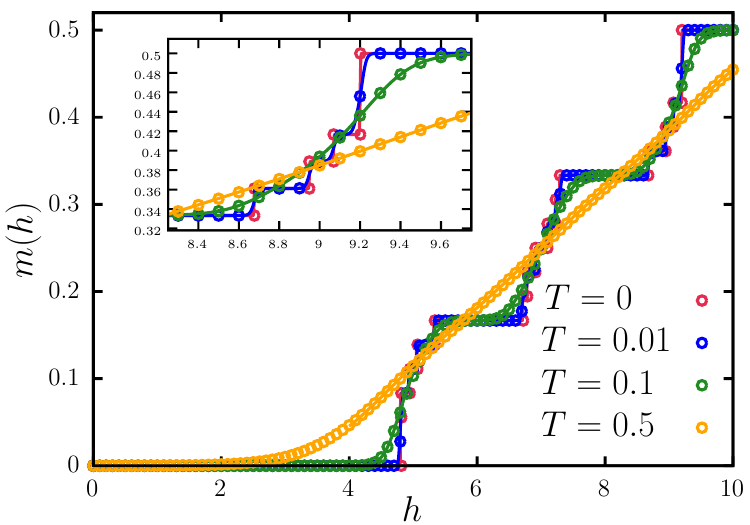}
\includegraphics[width=\columnwidth]{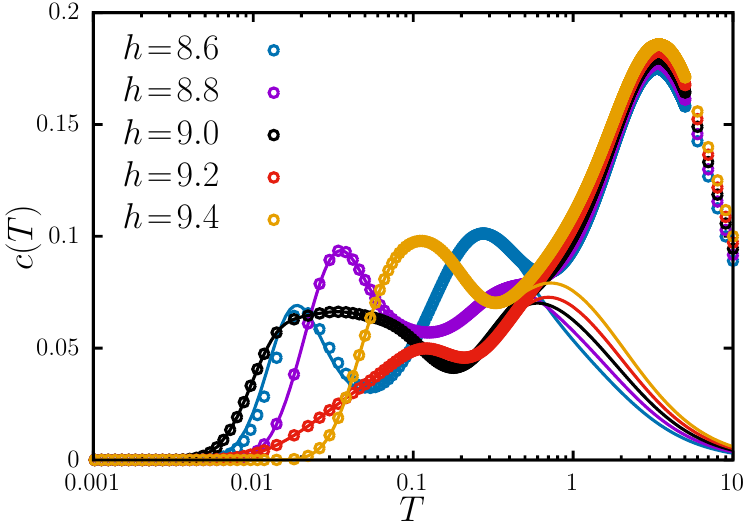}
\caption{Top: Magnetization curve $m(h)$ for the $S =1/2$ frustrated kagome-lattice bilayer with $J_1 = 1.1$, $J_{\sf x}=0.9$, $J_2 = 5$, $N = 36_a$; $T=0,\,0.01,\,0.1,\,0.5$. Inset is a zoom-in of the region around $h=9$.
Bottom: Specific heat $c(T)$ for the same system at $h= 8.6,\, 8.8,\, 9,\,9.2,\,9.4$.
In both panels along with numerics for the 36-site initial model (\ref{01}) (circles) we also show the results of the effective theory (\ref{20}) with ${\cal N} = 18_a$ (solid curves).}
\label{f09}
\end{figure}

In this section, we examine what happens around the fully frustrated case, when the ideal flat-band condition $J_1 = J_{\sf x}$ is slightly violated. 
Numerics for the magnetization curve and the temperature dependence of the specific heat for $J_1=1.1$, $J_{\sf x}=0.9$ are reported in Fig.~\ref{f09} (circles). 
Next, we discuss an effective theory explaining these data, see Refs.~\cite{Derzhko2013,Richter2018}.

In the regime when $J_2$ is the dominating coupling, one may elaborate a strong-coupling approach and obtain an effective theory which is valid for $J_1\ne J_{\sf x}$, too.
Within this approach, one starts from noninteracting singlets at the magnetic field $h_0=J_2$, which are governed by the Hamiltonian $H_{\rm main}=\sum_{{\bf m}\alpha}[J_2({\bf t}^2_{{\bf m}\alpha}-3/2)/2-h_0t_{{\bf m}\alpha}^z]$, cf. Eq.~(\ref{08}). The remaining Hamiltonian, i.e., $V=H-H_{\rm main}$, is treated as a perturbation. Clearly, the ground state of $H_{\rm main}$ is $2^{\cal N}$-fold degenerate and forms the model space defined by the projector $P=\otimes_{{\bf m}\alpha}(\vert u\rangle\langle u\vert + \vert d\rangle\langle d\vert)_{{\bf m}\alpha}$, where $\vert u\rangle=\vert\uparrow_a\uparrow_b\rangle$ and $\vert d\rangle=(\vert\uparrow_a\downarrow_b\rangle-\vert\downarrow_a\uparrow_b\rangle)/\sqrt{2}$. Switching on $V\ne 0$ lifts the degeneracy in the model space, and an effective Hamiltonian $H_{\rm eff}$, which acts in this space and describes the low-energy properties of $H$ around the limit $V=0$, can be constructed perturbatively \cite{Fulde1995}. Calculating the first term in the expansion $H_{\rm eff}=P(H_{\rm main}+V)P+{\cal{O}}(V^2)$ and introducing the pseudo-spin one-half operators $T^z=(\vert u\rangle\langle u\vert-\vert d\rangle\langle d\vert)/2$, $T^+=\vert u\rangle\langle d\vert$, $T^-=\vert d\rangle\langle u\vert$, one gets $P(t^x,t^y,t^z)P=(0,0,1/2+T^z)$, $P(d^x,d^y,d^z)P=(-\sqrt{2}T^x,-\sqrt{2}T^y,0)$ and finally arrives at the $S=1/2$ $XXZ$ model in a field on the kagome lattice. That is, the effective Hamiltonian reads:
\begin{eqnarray}
\label{20}
H_{\rm eff}
&=&
{\cal C}{\cal N}
-{\sf h}\sum_m T_m^z
\\
&&+\sum_{\langle mn\rangle}\left[
{\rm J}^zT_m^zT_m^z {+} {\sf J}\left(T_m^xT_n^x{+}T_m^yT_n^y\right)\right],
\nonumber\\
{\cal C}
&=&-\frac{h}{2}{-}\frac{J_2}{4}{+}\frac{J_1{+}J_{\sf x}}{4},
\;\;\;
{\sf h}=h{-}J_2{-}J_1{-}J_{\sf x},
\nonumber\\
{\rm J}^z
&=&
\frac{J_1{+}J_{\sf x}}{2},
\;\;\;
{\sf J}=J_1{-}J_{\sf x}
\ .
\nonumber
\end{eqnarray}
Here the first sum ($m={{\bf m}\alpha}$) runs over the vertices of the kagome lattice and the second sum ($\langle mn\rangle=\langle {\bf m}\alpha {\bf n}\beta\rangle$) runs over the edges of the kagome lattice.  
The ideal flat-band geometry $J_1-J_{\sf x}=0$ corresponds to the kagome-lattice Ising model (${\sf J}=0$) and is simply another representation of the lattice-gas model discussed above, cf. Eq.~(\ref{15}).  
In general, we arrive at the $S=1/2$ $XXZ$ model with easy axis anisotropy, $\vert {\sf J}\vert \ll {\sf J}^z$, ${\sf J}^z>0$, in a field on the kagome lattice, for which many results are known to date, see, e.g., Ref.~\cite{Ulaga2024} and references therein. 

The presented field and temperature dependencies in Fig.~\ref{f09} show good agreement between exact numerics ($N=36$) and effective-theory predictions (${\cal N}=18$) for the chosen set of parameters. Interestingly, effective model (\ref{20}) is constructed perturbatively around the main part $H_{\rm main}$ defined above, i.e., near the limit of noninteracting dimers at the degenerate singlet-triplet point ($J_2>0$, $J_1=J_{\sf x}=J=0$, $h=h_0=J_2$), and therefore $h-h_0$ and $J_1,$ $J_{\sf x}$ are implied to be small in Eq.~(\ref{20}). Besides, $T$ also cannot be too large, because of the number of states taking into account in Eq.~(\ref{20}): $2^{\cal N}$ instead of $2^N$, this becomes visible at $T\approx 0.5$ in Fig.~\ref{f09}, bottom. Nevertheless, the magnetization curve, Fig.~\ref{f09}, top, is reproduced by Eq.~(\ref{20}) quite well for all $h$ and $T=0\ldots0.5$. 
Again, the effective description has an advantage for examining the frustrated quantum spin model (\ref{01}): Although the spin model (\ref{20}) remains a frustrated quantum spin system, it contains a two times smaller number of sites, ${\cal N}=N/2$.

\section{Summary}
\label{s5}

The present study continues the examination of the $S=1/2$ Heisenberg model on frustrated bilayer lattices in the regime of strong antiferromagnetic interlayer coupling. In this paper, we studied the case of the kagome-lattice bilayer. Such a bilayer shows a prominent difference to other ones such as square-, honeycomb-, or triangular-lattice bilayers. While the latter bilayers for $h \lesssim h_{\rm sat}$ exhibit a ground-state ordering of singlets which persists up to a finite temperature $T_c(h)$, the former one does not show a symmetry breaking, but rather a huge degeneracy of the ground state, which grows exponentially with lattice size. This degeneracy has been known in combinatorics for many years. 
Two features of the kagome-lattice bilayer at low temperatures and just below the saturation field are peculiar and interesting: There is no phase transition related to the singlets ordering, and the bulk specific heat may depend on the sample form.

Concerning the experimental side, one may mention several quantum Heisenberg antiferromagnets on frustrated bilayer lattices: 
Ba$_2$CoSi$_2$O$_6$Cl$_2$ \cite{Tanaka2014,Kurita2019} is a square-lattice bilayer,
Bi$_3$Mn$_4$O$_{12}$(NO$_3$) \cite{Matsuda2010,Kandpal2011} is a honeycomb-lattice bilayer,
and
K$_2$Co$_2$(SeO$_3$)$_3$ \cite{Fu2025} is a triangular-lattice bilayer.
Unfortunately, we are not aware of a solid-state realization of the frustrated kagome-lattice bilayer quantum Heisenberg antiferromagnet yet.

Finally, let us mention several problems which in our opinion deserve further studies.
As follows from the calculation of the entropy as a function of temperature and magnetic field (Sec.~\ref{s3c}), the system at hand should exhibit an interesting magnetocaloric effect since the residual entropy remains a nonzero constant for the 2/3- and 1/3-magnetization plateaus \cite{Zhitomirsky2003,Zhu2003,Schnack2020b}.
In view of Elser's formula (\ref{18}), see Sec.~\ref{s3d}, it is worth to examine the thermodynamics of the hexagonal-shaped systems in detail. To this end, one has to consider larger systems and adapt Monte-Carlo simulations accordingly. 
One may also consider other Archimedean lattice bilayers \cite{Yu2015,Farnell2018}, which, however, might not provide new physics in comparison to the cases studied already in Refs.~\cite{Derzhko2010,Krokhmalskii2017,Strecka2018} and in the present paper. 

\section*{Acknowledgements}

The Ukrainian scholars thank the Armed Forces of Ukraine for protecting them.
D.~Y. and T.~H. are grateful for the fellowships of the President of Ukraine for young scholars. T.~H. was supported by the Projects of research works of young scientists of the National Academy of Sciences of Ukraine (project \# 29-04/18-2023). 
T.~H., T.~K., and O.~D. were supported through the EURIZON project (\#3025 ``Frustrated quantum spin models to explain the properties of magnets over wide temperature range''), which is funded by the European Union under grant agreement No.~871072.
O.~D. acknowledges the kind hospitality of the University of Bielefeld in October-December of 2023 (supported by Erasmus+ and DFG). 
O.~D. acknowledges the hospitality of the MPIPKS, Dresden at the Intercontinental Binodal Workshop Flat bands and high-order Van Hove singularities (27 May - 7 June 2024)
and
thanks A.~Honecker and K.~Kar\ifmmode \check{l}\else \v{l}\fi{}ov\'a for their hospitality at the Advanced quantum materials for magnetic cooling and quantum information science conference (Cergy-Pontoise, France, 19-21 February 2025).
J.~S. thanks the Max Planck Institute for the Physics of Complex Systems for the hospitality during a visit in April 2025.
This work was supported by the Deutsche Forschungsgemeinschaft (DFG SCHN 615/28-1 and RI 615/25-1).

\appendix

\section{Numerics for small kagome bilayers}
\label{aa}
\renewcommand{\theequation}{A.\arabic{equation}}
\setcounter{equation}{0}

\begin{figure}
\includegraphics[width=\columnwidth]{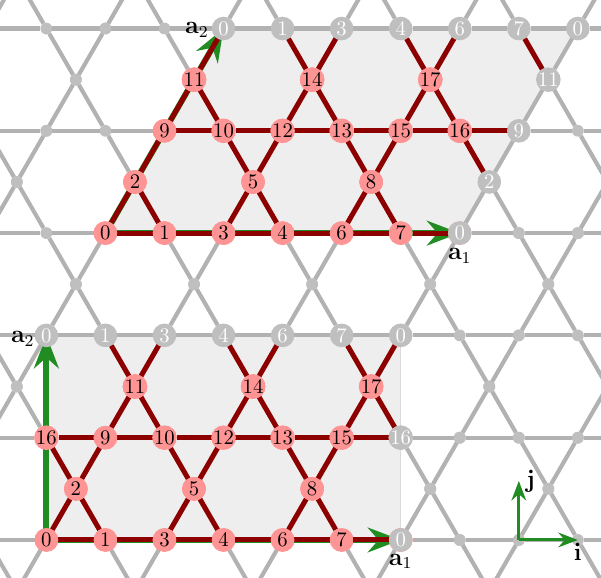} 
\caption{Two types of periodic boundary conditions for $N=36$ (${\cal N}=18$), 
i.e., 
(top) determined by the edge vectors ${\bf a}_1=(6,0)$, ${\bf a}_2=(2,2\sqrt{3})$ ($N=36a$ or ${\cal N}=18a$) 
and 
(bottom) determined by the edge vectors ${\bf a}_1=(6,0)$, ${\bf a}_2=(0,2\sqrt{3})$ ($N=36b$ or ${\cal N}=18b$).}
\label{f10}
\end{figure}

In this appendix, we present more results obtained numerically for small lattices.
In Fig.~\ref{f10} we illustrate two types of the 36-site frustrated kagome-lattice bilayer which are used for numerical calculations in our study, cf. Table~\ref{t03} and Table~\ref{t04}.

\begin{table}
\caption{\label{t03}
Numerics for the kagome-lattice bilayer of $N=36_a$ sites [periodic boundary conditions, the edge vectors are ${\bf a}_1=(6,0)$ and ${\bf a}_2=(2,2\sqrt{3})$], $J=1$, and (from top to bottom) $J_2=5$, $3$, $2$, and $1$.
We present the energy $E_{\rm GS}$ and degeneracy ${\cal W}_{\rm GS}$ of the ground state in various subspaces $12\le S^z\le 17$. We also present the energy ${E_1}_{\rm min}$ and degeneracy ${{\cal W}_1}_{\rm min}$ of the first excited state as well as the energy gap ${E_1}_{\rm min}{-}E_{\rm GS}$ in  various subspaces $12\le S^z\le 17$.}
\begin{ruledtabular}
\begin{tabular}{llllll}
$S^z$ & $E_{\rm GS}$ & ${\cal W}_{\rm GS}$ & ${E_1}_{\rm min}$ & ${{\cal W}_1}_{\rm min}$ & ${E_1}_{\rm min}{-}E_{\rm GS}$ \\
\hline
17 & 49.5 &  18 & 52.5 &    7 & 3 \\
16 & 40.5 & 117 & 41.5 &   36 & 1 \\
15 & 31.5 & 336 & 32.5 &  396 & 1 \\     
14 & 22.5 & 417 & 23.5 & 1356 & 1 \\
13 & 13.5 & 186 & 14.5 & 1572 & 1 \\
12 &  4.5 &  17 &  5.5 &  456 & 1 \\
\hline                        
17 & 42.5 &  18 & 43.5 &    7 & 1 \\      
16 & 35.5 & 117 & 36.5 &  108 & 1 \\      
15 & 28.5 & 336 & 29.5 &  585 & 1 \\ 
14 & 21.5 & 417 & 22.5 & 1456 & 1 \\      
13 & 14.5 & 186 & 15.5 & 1584 & 1 \\              
\hline
17 &   39 &  25 & 40.267\,9\ldots &  2 & 1.267\,9\ldots \\     
16 &   33 & 199 & 33.279\,0\ldots &  3 & 0.279\,0\ldots \\ 
15 &   27 & 544 & 27.063\,6\ldots & 12 & 0.063\,6\ldots \\ 
14 &   21 & 523 & 21.106\,9\ldots & 24 & 0.106\,9\ldots \\ 
13 &   15 & 198 & 15.224\,4\ldots & 48 & 0.224\,4\ldots \\       
12 &    9 &  17 &  9.620\,8\ldots & 24 & 0.620\,8\ldots \\ 
\hline
17 & 34.5            &  7 & 35.5            & 18 & 1              \\ 
16 & 28.5            & 10 & 28.779\,0\ldots &  3 & 0.279\,0\ldots \\      
15 & 22.5            &  1 & 22.650\,9\ldots &  1 & 0.150\,9\ldots \\      
14 & 17.028\,5\ldots &  1 & 17.260\,8\ldots &  3 & 0.232\,3\ldots \\ 
13 & 11.950\,5\ldots & 1  & 12.185\,4\ldots &  3 & 0.234\,9\ldots \\ 
\end{tabular}
\end{ruledtabular}
\end{table}

\begin{table}
\caption{\label{t04}
Numerics for the kagome-lattice bilayer of $N=36_b$ sites 
[periodic boundary conditions, the edge vectors ${\bf a}_1=(6,0)$ and ${\bf a}_2=(0,2\sqrt{3})$], $J=1$, and (from top to bottom) $J_2=5$, $3$, $2$, and $1$. We present the same quantities as in Table~\ref{t03}.}
\begin{ruledtabular}
\begin{tabular}{llllll}
$S^z$ & $E_{\rm GS}$ & ${\cal W}_{\rm GS}$ & ${E_1}_{\rm min}$ & ${{\cal W}_1}_{\rm min}$ & ${E_1}_{\rm min}{-}E_{\rm GS}$ \\
\hline
17 & 49.5 &  18 & 52.5 &    7 & 3 \\
16 & 40.5 & 117 & 41.5 &   36 & 1 \\
15 & 31.5 & 336 & 32.5 &  396 & 1 \\     
14 & 22.5 & 420 & 23.5 & 1344 & 1 \\
13 & 13.5 & 192 & 14.5 & 1560 & 1 \\
12 &  4.5 &  20 &  5.5 &  456 & 1 \\
\hline                        
17 & 42.5 &  18 & 43.5 &    7 & 1 \\
16 & 35.5 & 117 & 36.5 &  108 & 1 \\
15 & 28.5 & 336 & 29.5 &  594 & 1 \\
14 & 21.5 & 420 & 22.5 & 1504 & 1 \\ 
13 & 14.5 & 192 & 15.5 & 1590 & 1 \\         
12 &  7.5 &  20 &  8.5 &  456 & 1 \\
\hline
17 &   39 &  25 & 40              &  2 & 1              \\    
16 &   33 & 199 & 33.198\,0\ldots &  7 & 0.198\,0\ldots \\
15 &   27 & 566 & 27.092\,6\ldots & 24 & 0.092\,6\ldots \\
14 &   21 & 595 & 21.117\,0\ldots & 24 & 0.117\,0\ldots \\
13 &   15 & 222 & 15.215\,9\ldots & 12 & 0.215\,9\ldots \\        
12 &    9 &  20 &  9.585\,7\ldots &  6 & 0.585\,7\ldots \\
\hline
17 & 34.5            &  7 & 35.5            & 20 & 1              \\
16 & 28.5            & 10 & 28.698\,0\ldots &  1 & 0.198\,0\ldots \\
15 & 22.5            &  2 & 22.678\,4\ldots &  1 & 0.178\,4\ldots \\
14 & 16.939\,9\ldots &  1 & 17.160\,9\ldots &  2 & 0.220\,9\ldots \\
13 & 11.995\,8\ldots &  1 & 12.064\,9\ldots &  1 & 0.069\,0\ldots \\
\end{tabular}
\end{ruledtabular}
\end{table}

In Tables~\ref{t03} and \ref{t04} we report some characteristics of the ground state and low-lying states (energy and degeneracy) in the subspaces with total $S^z=N/2,\ldots,N/3$ ($N=36$, two periodic clusters are examined) depending on $J_2$ to illustrate the emergence of the large-$J_2$ regime.
If $J_2$ exceeds $2J$ the low-energy states characteristics are well structured (easy understandable energy and degeneracy) and they do not depend on $J_2$. Namely, $E_{\rm GS}$ fits to $E_{\rm FM}+n\epsilon_0$ with $E_{\rm FM}=(2J+J_2/4){\cal N}$ and $\epsilon_0=-(4J+J_2)$ whereas $W_{\rm GS}$ for $J_2=5$ and $J_2=3$ fits to the number of spatial configurations of $n=1,\ldots,6$ hard rhombi on the kagome lattice of ${\cal N}=18$ sites, see Table~\ref{t01}. 
Besides, the ground-state degeneracy ${\cal W}_{\rm GS}$ for $S^z=12$ illustrates the degeneracy of the 2/3 plateau state  shown in Fig.~\ref{f02}. As can be seen in Tables~\ref{t03} and \ref{t04}, its value for $N=36$ depends on the cluster shape and is rather large being 17 (Table~\ref{t03}) or 20 (Table~\ref{t04}). For $N=42$ the ground-state degeneracy for $S^z=14$ is 31, i.e., it grows as $N$ increases. $J_2=2J$ is a marginal case, exhibiting higher degeneracy of the ground state: The flat-band states (\ref{06}) and (\ref{07}) have the same energy. Smaller $J_2$ indicate completely different nature of the ground state: No regularly structured energy and no large degeneracy are observed in that case.

The picture of overlapping rhombi is also confirmed by numerical data reported in Tables~\ref{t03} and \ref{t04}. For example, for $J_2=5$ the energy gap in the subspaces with $2\le n\le 6$ is $J=1$ and ${\cal W}_{1{\rm min}}$, according to the soft-rhombi calculations in Appendix~\ref{ab}, corresponds to the number of ways to put one pair of neighboring singlets while the rest singlets are not neighbors (patterns with one pair of overlapping rhombi, see the lower panel in Fig.~\ref{f04}).
  
\section{Rhombi on the kagome lattice}
\label{ab}
\renewcommand{\theequation}{B.\arabic{equation}}
\setcounter{equation}{0}

In this appendix, we provide more results about rhombi on the kagome lattice which complement those that have been reported in Sec.~\ref{s3}.

\begin{figure*}
\includegraphics[width=1.95\columnwidth]{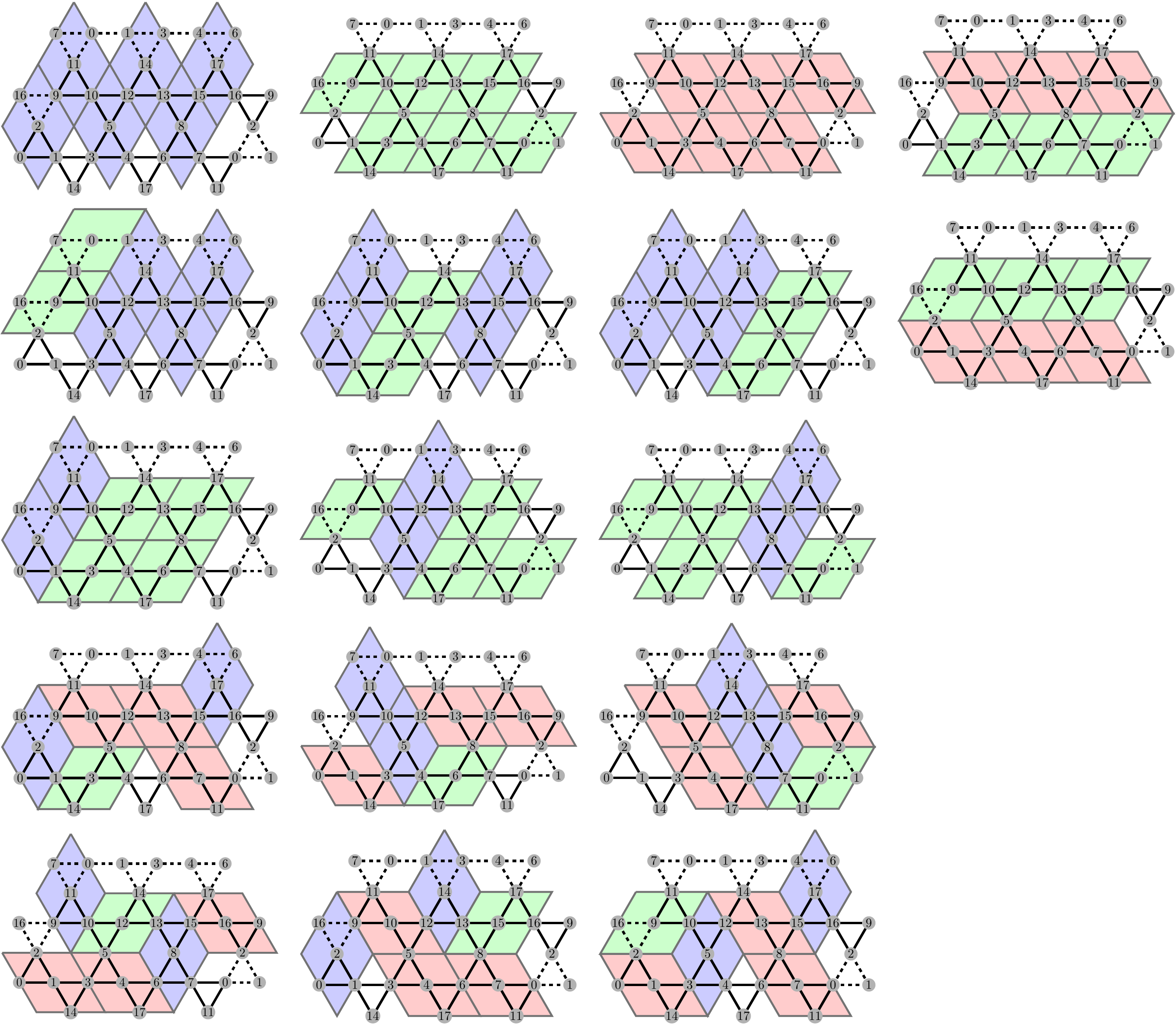}
\caption{17 hard-rhombi coverings of the periodic kagome lattice of ${\cal N}=18_a$ sites. Numerics for the periodic spin system $N=36_a$ gives for the localized-magnon crystal state the same degeneracy ${\cal W}_{18}=17$, see Table~\ref{t03}.}
\label{f11}
\end{figure*}

\begin{figure*}	
\includegraphics[width=1.95\columnwidth]{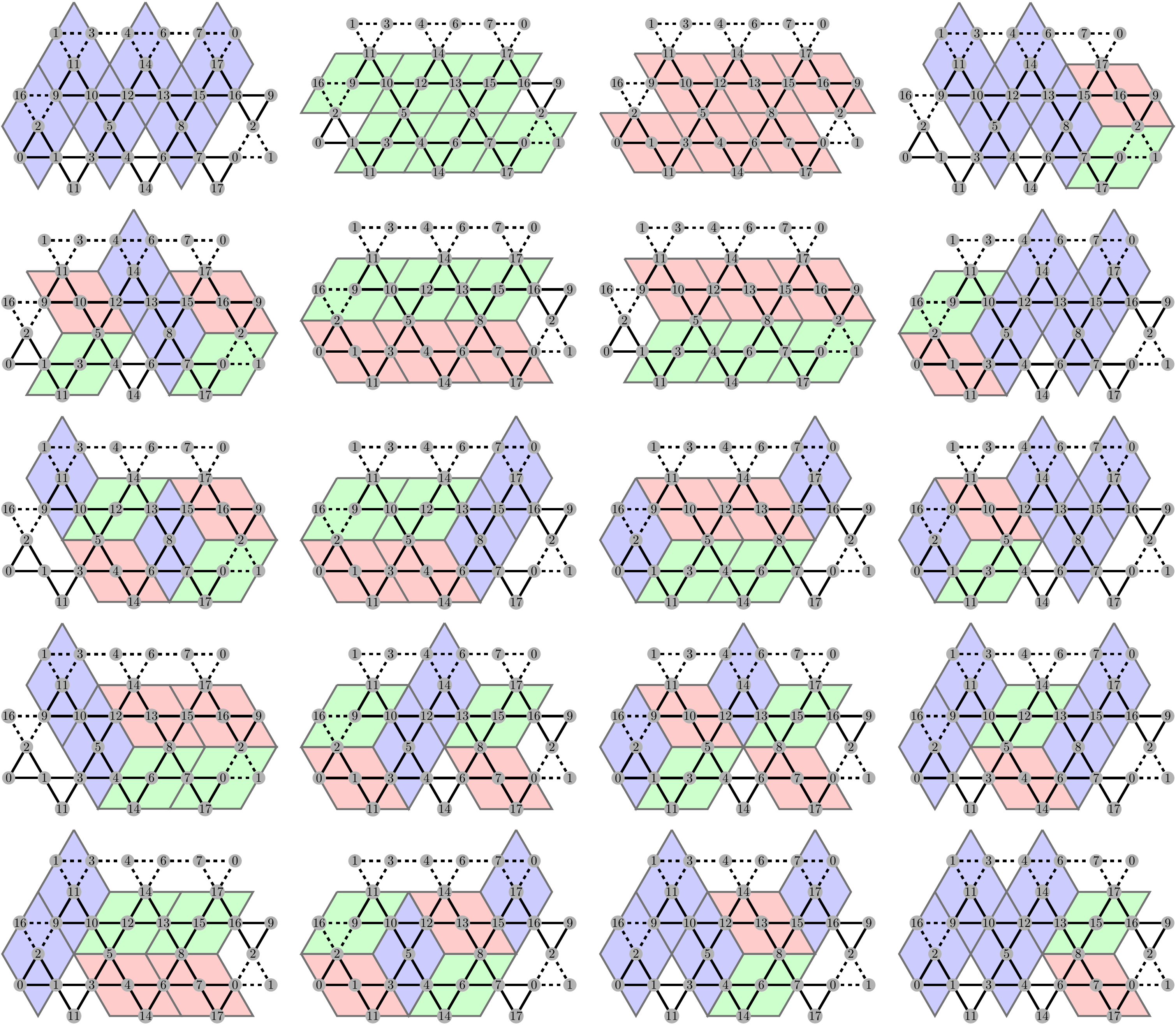}
\caption{20 hard-rhombi coverings of the periodic kagome lattice of ${\cal N}=18_b$ sites. Numerics for the periodic spin system $N=36_b$ gives for the localized-magnon crystal state the same degeneracy ${\cal W}_{18}=20$, see Table~\ref{t04}.}
\label{f12}
\end{figure*}

First, in Figs.~\ref{f11} and \ref{f12} we illustrate all spatial configurations of $n_{\max}=6$ hard rhombi on the ${\cal N}=18$ kagome lattice with two types of periodic boundary conditions, ${\cal N}=18_a$ and ${\cal N}=18_b$, see Fig.~\ref{f10}. These configurations may be found by direct search. Furthermore, they may be encoded by indicating the sites which are occupied by rhombi.
Thus, in Fig.~\ref{f11} we have (from left to right beginning from the top):
$2,5,8,11,14,17$;
$0,3,6,9,12,15$;
$1,4,7,10,13,16$;
$0,3,6,10,13,16$;
$0,5,8,9,14,17$;
$2,3,8,11,12,17$;
$2,5,6,11,14,15$;
$1,4,7,9,12,15$;
$2,3,6,11,12,15$;
$0,5,6,9,14,15$;
$0,3,8,9,12,17$;
$2,3,7,10,13,17$;
$1,5,6,11,13,16$;
$0,4,8,10,14,16$;
$1,4,8,11,12,16$;
$2,4,7,10,14,15$;
and
$1,5,7,9,13,17$.
Similarly, in Fig.~\ref{f12} we have (from left to right beginning from the top):
$2,5,8,11,14,17$;
$0,3,6,9,12,15$;
$1,4,7,10,13,16$;
$0,5,8,11,14,16$;
$0,3,8,10,14,16$;
$1,4,7,9,12,15$;
$0,3,6,10,13,16$;
$1,5,8,9,14,17$;
$0,4,8,11,12,16$;
$1,4,8,9,12,17$;
$2,3,6,10,13,17$;
$2,3,8,10,14,17$;
$0,5,6,11,13,16$;
$1,5,7,9,14,15$;
$2,3,7,10,14,15$;
$2,4,8,11,12,17$;
$2,4,7,11,12,15$;
$1,5,6,9,13,17$;
$2,5,6,11,13,17$;
and
$2,5,7,11,14,15$.

Next, we explain how we systematically count the spatial configurations of hard rhombi. A necessary condition to get an allowed spatial configuration is to put only one rhombus per unit cell. There are ${\cal N}/3$ unit cells, therefore, in the case of $1< n\le {\cal N}/3$ rhombi, we consider as the first step ${\cal C}_{{\cal N}/3}^n$ spatial configurations for which each rhombus is assigned to a specific unit cell. Since there are 3 sites in the unit cell, we consider as the second step $3^n$ ways to rearrange rhombus inside the unit cells. For each obtained in such a way spatial configuration of $n$ rhombi on the ${\cal N}$-site kagome lattice we verify the hard-core rule: If it holds, the spatial configuration is accepted. 
Along these lines we obtain $g_{\cal N}(n)$ reported in Tables~\ref{t01} and \ref{t02}. In Tables~\ref{t05} and \ref{t06} we report the results similar to what is given in Table~\ref{t01}, but for other lattice sizes ${\cal N}=21$ and ${\cal N}=12,\,24,\,27,\,36$. 

\begin{table}
\caption{\label{t05}
		Counting the number of $n$ hard-rhombi spatial configurations on the kagome lattice of ${\cal N}=21$ sites.}
\begin{ruledtabular}
\vspace{3mm}
\begin{tabular}{r||r}
			~$n$~ & $g_{21}(n)$ \\
			\hline \hline
			~1~ &   ~21~  \\ 
			~2~ &  ~168~  \\
			~3~ &  ~644~  \\
			~4~ & ~1~225~ \\
			~5~ & ~1~085~ \\
			~6~ &   ~371~ \\
			~7~ &    ~31~ \\
\end{tabular}          
\end{ruledtabular}
\end{table}

\begin{table}
\caption{\label{t06}
Counting the number of $n$ hard-rhombi spatial configurations on the kagome lattise of ${\cal N}$ sites with two types of periodic boundary conditions imposed. These data are complementary to the ones in Table~\ref{t01}.}
\begin{ruledtabular}
\vspace{1mm}
${\cal{N}}=12$
\vspace{1mm}
\begin{tabular}{r||rr}
~$n$~ & $g_{12_a}(n)$, ~${\cal N}_a$~ & $g_{12_b}(n)$, ~${\cal N}_b$~ \\
\hline \hline
~1~ & ~12~ & ~12~ \\
~2~ & ~42~ & ~42~ \\
~3~ & ~44~ & ~44~ \\
~4~ &  ~9~ &  ~9~ \\
\end{tabular}
\vspace{3mm}
${\cal{N}}=24$
\vspace{1mm}
\begin{tabular}{r||rr}
~$n$~ & $g_{24_a}(n)$, ~${\cal N}_a$~ & $g_{24_b}(n)$, ~${\cal N}_b$~ \\
\hline \hline
~1~ &    ~24~ &    ~24~ \\ 
~2~ &   ~228~ &   ~228~ \\
~3~ & ~1~096~ & ~1~096~ \\
~4~ & ~2~830~ & ~2~834~ \\
~5~ & ~3~848~ & ~3~880~ \\
~6~ & ~2~516~ & ~2~588~ \\
~7~ &   ~632~ &   ~696~ \\
~8~ &    ~33~ &    ~49~ \\
\end{tabular}
\vspace{3mm}
${\cal{N}}=27$
\vspace{1mm}
\begin{tabular}{r||rr}
~$n$~ & $g_{27_a}(n)$, ~${\cal N}_a$~ & $g_{27_b}(n)$, ~${\cal N}_b$~ \\
\hline \hline
~1~ &     ~27~ &     ~27~ \\ 
~2~ &    ~297~ &    ~297~ \\
~3~ &  ~1~719~ &  ~1~719~ \\
~4~ &  ~5~643~ &  ~5~643~ \\
~5~ & ~10~557~ & ~10~557~ \\
~6~ & ~10~737~ & ~10~740~ \\
~7~ &  ~5~319~ &  ~5~328~ \\
~8~ &  ~1~026~ &  ~1~035~ \\
~9~ &     ~42~ &     ~45~ \\
\end{tabular}
\vspace{3mm}
${\cal{N}}=36$
\vspace{1mm}
\begin{tabular}{r||rr}
~$n$~ & $g_{36_a}(n)$, ~${\cal N}_a$~ & $g_{36_b}(n)$, ~${\cal N}_b$~ \\
\hline \hline
~1~ &      ~36~ &      ~36~ \\ 
~2~ &     ~558~ &     ~558~ \\
~3~ &   ~4~884~ &   ~4~884~ \\
~4~ &  ~26~613~ &  ~26~613~ \\
~5~ &  ~93~888~ &  ~93~888~ \\
~6~ & ~216~232~ & ~216~240~ \\
~7~ & ~320~304~ & ~320~400~ \\
~8~ & ~293~637~ & ~294~045~ \\
~9~ & ~155~364~ & ~156~116~ \\
~10~ &  ~42~006~ &  ~42~606~ \\
~11~ &   ~4~596~ &   ~4~788~ \\
~12~ &     ~113~ &     ~129~ \\
\end{tabular}
\end{ruledtabular}
\end{table}

\begin{widetext}

\begin{table}
\caption{\label{t07}
Nonzero ${\cal G}(\Sigma_1,\Sigma_2)$ for the kagome lattice of ${\cal N}=18_a$ sites, see Eq.~(\ref{b01}). Along the row: $\Sigma_1=0,1,\ldots18$; along the column: $\Sigma_2=0,1,\ldots,36$.}
\begin{ruledtabular}
\vspace{3mm}
\begin{tabular}{r||r|r|r|r|r|r|r|r|r|r|r|r|r|r|r|r|r|r|r|}
				& 0 & 1 & 2 & 3 & 4 & 5 & 6 & 7 & 8 & 9 & 10 & 11 & 12 & 13 & 14 & 15 & 16 & 17 & 18\\
				\hline \hline
				0 & 1 & 18 & 117& 336 & 417 & 186 & 17 & & & & & & & & & & & &  \\ 
				1 & & & 36 & 396 & 1\,356 & 1\,572 & 456 & & & & & & & & & & & & \\
				2 & & & & 72 & 972 & 3\,402 & 3\,192 & 444 & & & & & & & & & & & \\
				3 & & & & 12 & 240 & 2\,160 & 6\,104 & 3\,588 & & & & & & & & & & & \\
				4 & & & & & 75 & 966 & 5\,001 & 8\,712 & 2\,190 & & & & & & & & & & \\
				5 & & & & & & 252 & 2\,604 & 8\,712 & 7\,548 & & & & & & & & & & \\
				6 & & & & & & 30 & 1\,004 & 6\,672 & 12\,048 & 3\,580 & & & & & & & & & \\
				7 & & & & & & & 168 & 2\,544 & 11\,256 & 8\,400 & & & & & & & & & \\
				8 & & & & & & & 18 & 1\,032 & 6\,849 & 14\,160 & 2\,190 & & & & & & & & \\
				9 & & & & & & & & 108 & 2\,880 & 11\,400 & 7\,548 & & & & & & & & \\
				10 & & & & & & & & 12 & 864 & 7\,284 & 12\,048 & 444 & & & & & & & \\
				11 & & & & & & & & & 120 & 2\,832 & 11\,256 & 3\,588 & & & & & & & \\  
				12 & & & & & & & & & 3 & 844 & 6\,849 & 8\,712 & 17 & & & & & & \\  
				13 & & & & & & & & & & 120 & 2\,880 & 8\,712 & 456 & & & & & & \\ 
				14 & & & & & & & & & & & 864 & 6\,672 & 3\,192 & & & & & & \\ 
				15 & & & & & & & & & & & 120 & 2\,544 & 6\,104 & & & & & & \\ 
				16 & & & & & & & & & & & 3 & 1\,032 & 5\,001 & 186 & & & & & \\
				17 & & & & & & & & & & & & 108 & 2\,604 & 1\,572 & & & & & \\
				18 & & & & & & & & & & & & 12 & 1\,004 & 3\,402 & & & & & \\ 
				19 & & & & & & & & & & & & & 168 & 2\,160 & & & & & \\ 
				20 & & & & & & & & & & & & & 18 & 966 & 417 & & & & \\ 
				21 & & & & & & & & & & & & & & 252 & 1\,356 & & & & \\ 
				22 & & & & & & & & & & & & & & 30 & 972 & & & & \\ 
				23 & & & & & & & & & & & & & & & 240 & & & & \\
				24 & & & & & & & & & & & & & & & 75 & 336 & & & \\  
				25 & & & & & & & & & & & & & & & & 396 & & & \\
				26 & & & & & & & & & & & & & & & & 72 & & & \\
				27 & & & & & & & & & & & & & & & & 12 & & & \\
				28 & & & & & & & & & & & & & & & & & 117 & & \\
				29 & & & & & & & & & & & & & & & & & 36 & & \\
				30 & & & & & & & & & & & & & & & & & & & \\
				31 & & & & & & & & & & & & & & & & & & & \\
				32 & & & & & & & & & & & & & & & & & & 18 & \\
				33 & & & & & & & & & & & & & & & & & & & \\
				34 & & & & & & & & & & & & & & & & & & & \\
				35 & & & & & & & & & & & & & & & & & & & \\
				36 & & & & & & & & & & & & & & & & & & & 1 \\
\end{tabular}          
\end{ruledtabular}
\end{table}

\end{widetext}

Finally, we explain how we work with the soft rhombi configurations. It is convenient to cast Eq.~(\ref{13}) into
\begin{eqnarray}
\label{b01}
\Xi(z,{\cal N})=\sum_{\Sigma_1=0}^{{\cal N}}\sum_{\Sigma_2=0}^{2{\cal N}}{\cal G}(\Sigma_1,\Sigma_2)z^{\Sigma_1}{\rm e}^{-\frac{V\Sigma_2}{T}}.
\end{eqnarray}
To arrive at Eq.~(\ref{b01}), we denote $n_1+\ldots+n_{\cal N}=\Sigma_1$ and note that $\Sigma_1$ may acquire the values $0,1,\ldots,{\cal N}$, denote $\sum_{\langle ij\rangle}n_in_j=\Sigma_2$ and note that $\Sigma_2$ may acquire, in principle, the values $0,1,\ldots,2{\cal N}$, and introduce ${\cal G}(\Sigma_1,\Sigma_2)$ which stands for the number of spatial configurations of $n=0,1,\ldots,{\cal N}$ soft rhombi which yields $\Sigma_1=n$ and $\Sigma_2$ [${\cal G}(\Sigma_1,\Sigma_2)$ may be zero for some values of $\Sigma_1,\Sigma_2$]. Clearly, $\sum_{\Sigma_1=0}^{{\cal N}}\sum_{\Sigma_2=0}^{2{\cal N}}{\cal G}(\Sigma_1,\Sigma_2)=2^{\cal N}$. 
In the hard-rhombi limit only the terms with $\Sigma_2=0$ in Eq.~(\ref{b01}) survive, $\Sigma_1=n$ and ${\cal G}(\Sigma_1,0)=g_{\cal N}(n)$. Furthermore, the number of patterns with one pair of overlapping rhombi is given by ${\cal G}(\Sigma_1,1)$, $\Sigma_1\ge 2$ (cf. the corresponding row in Table~\ref{t07} and ${\cal W}_{1\min}=36,\,396,\,1\,356,\,1\,572,\,456$ in Table~\ref{t03}). 
Finding ${\cal G}(\Sigma_1,\Sigma_2)$ is the most difficult task. For that we have to find the values of $\Sigma_1$ and $\Sigma_2$ for all $2^{\cal N}$ spatial configurations of $n=0,\ldots {\cal N}$ soft rhombi on the ${\cal N}$-site kagome lattice. In Table~\ref{t07} we report the output of such calculations for the case ${\cal N}=18_a$ for illustration. The reported numbers satisfy $\sum_{\Sigma_1=0}^{18}\sum_{\Sigma_2=0}^{36}{\cal G}(\Sigma_1,\Sigma_2)=2^{18}$. They also illustrate a ``particle-hole'' symmetry: Compare the columns for $\Sigma_1=0,\ldots,8$ and for $\Sigma_1=18,\ldots,10$ in Table~\ref{t07}. 

\newpage

\bibliography{ka_bi_refs_1}

\end{document}